\documentclass[11pt]{article}

\usepackage{caption}
\usepackage{amssymb}
\usepackage[margin=1.25in]{geometry}
\usepackage{lineno}
\usepackage{rotating}
\usepackage{multirow}
\usepackage{booktabs}
\usepackage[dvipsnames,usenames]{color}
\RequirePackage{url} 
\usepackage[symbol]{footmisc}   
\usepackage{booktabs}     

\begin{document}

%
\newcounter{fig}
\newcommand{\mylist}[2]
  {\begin{list}{#1}{
      \setlength{\topsep}{0pt}
      \setlength{\parsep}{0.0in}
      \setlength{\itemsep}{#2}
  }}
\newcommand{\newlist}[3]
  {\begin{list}{#1}{
      \setlength{\topsep}{0pt}
      \setlength{\parsep}{0.0in}
      \setlength{\itemsep}{#2}
      \setlength{\leftmargin}{#3}
  }}
\newcommand{\newenum}[1]
  {
  \begin{list}{\arabic{fig}.}{\usecounter{fig}
      \setlength{\topsep}{0pt}
      \setlength{\parsep}{0.0in}
      \setlength{\itemsep}{#1}
  }}
%
%
\newcommand{\newenumr}[1]
  {
  \begin{list}{\Roman{fig}.}{\usecounter{fig}
      \setlength{\topsep}{0 pt}
      \setlength{\parsep}{0.0in}
      \setlength{\itemsep}{#1}
  }}
%
%
\newenvironment{mynum}{
\begin{enumerate}
      \setlength{\itemsep}{1 pt}
      \setlength{\parskip}{0 pt}
      \setlength{\parsep}{0 pt}
      \setlength{\topsep}{-20 pt}
      \setlength{\partopsep}{-20 pt}
}
{\end{enumerate}}
\newcommand{\newnum}
  {\begin{enumerate}{
      \setlength{\itemsep}{0 pt}
      \setlength{\parskip}{0 pt}
      \setlength{\parsep}{0 pt}
      \setlength{\topsep}{-20 pt}
      \setlength{\partopsep}{-20 pt}
  }}
\newenvironment{changemargin}[2]{%
 \begin{list}{}{%
  \setlength{\topsep}{0pt}%
  \setlength{\leftmargin}{#1}%
  \setlength{\rightmargin}{#2}%
  \setlength{\listparindent}{\parindent}%
  \setlength{\itemindent}{\parindent}%
  \setlength{\parsep}{\parskip}%
 }%
\item[]}{\end{list}}
%
%
\let\la=\lessim
\newcommand{\D}{\displaystyle}
\newcommand{\T}{\textstyle}
\newcommand{\bvec}[1]{\vec{\mathbf{#1}}}
\newcommand{\bhat}[1]{\hat{\mathbf{#1}}}
\newcommand{\gam}{\mbox{$\Gamma$}}
\newcommand{\agam}{\mbox{$\overline{\Gamma}$}}
\newcommand{\cp}{\mbox{\it CP}}
\newcommand{\ra}{\mbox{~$\rightarrow$}~} 
\newcommand{\nra}{\mbox{$\mid\rightarrow$}~} 
\newcommand{\tra}{
  \setlength{\unitlength}{1ex}
  \begin{picture}(0,2)
    \put(0.1,.6){\line(0,1){1.4}}
    \end{picture}$\rightarrow$~
  }
\newcommand{\la}{\mbox{~$\leftarrow$}~} 
\newcommand{\lra}{\mbox{~$\leftrightarrow$}~}
\newcommand{\llra}{\mbox{~$\Longleftrightarrow$}~}
\newcommand{\ccc}{\mbox{$^\circ$}}
\newcommand{\tc}{\mbox{$^\circ$C}}
\newcommand{\crc}{\mbox{$^{\circ}$}}
\newcommand{\mss}{\mbox{m/s$^2$}}
\newcommand{\kmc}{\mbox{kg/m$^3$}}
\newcommand{\cd}{\mbox{$\cdot$}}
\newcommand{\inch}{\mbox{$^{\prime\prime}$}}
\newcommand{\ft}{\mbox{$^{\prime}$}}
\newcommand{\gevc}{\mbox{GeV/$c$}}
\newcommand{\gevcc}{\mbox{GeV/$c^2$}}
%
\newcommand{\hypercp}
           {\mbox{\slshape Hyper}\raisebox{-.2ex}{\slshape C}%
                                 \raisebox{.2ex}{\slshape P}}
\newcommand{\dz}{\mbox{D\O}}
\newcommand{\minerva}{\mbox{MINER$\nu$A}}
\newcommand{\mute}{\mbox{Mu2e}}
\newcommand{\nova}{\mbox{NOvA}}
\newcommand{\pwo}{\mbox{PbWO$_4$}}
%
%
%
\newcommand{\gNuc}{\mbox{\it N}}
%
\newcommand{\gG}{\mbox{\it g}}
%
\newcommand{\gW}{\mbox{\it W}}
%
\newcommand{\gH}{\mbox{\it H}}
%
\newcommand{\gq}{\mbox{\it q}}
\newcommand{\uq}{\mbox{\it u}}
\newcommand{\dq}{\mbox{\it d}}
\newcommand{\cq}{\mbox{\it c}}
\newcommand{\sq}{\mbox{\it s}}
\newcommand{\tq}{\mbox{\it t}}
\newcommand{\bq}{\mbox{\it b}}
%
\newcommand{\agq}{\mbox{$\bar{q}$}}
\newcommand{\auq}{\mbox{$\bar{u}$}}
\newcommand{\adq}{\mbox{$\bar{d}$}}
\newcommand{\acq}{\mbox{$\bar{c}$}}
\newcommand{\asq}{\mbox{$\bar{s}$}}
\newcommand{\atq}{\mbox{$\bar{t}$}}
\newcommand{\abq}{\mbox{$\bar{b}$}}
%
\newcommand{\X}{\mbox{\it X}}
\newcommand{\Xm}{\mbox{\it X$^-$}}
\newcommand{\Xp}{\mbox{\it X$^+$}}
\newcommand{\Xz}{\mbox{\it X$^{\circ}$}}
%
\newcommand{\gl}{\mbox{\it l}}
\newcommand{\lm}{\mbox{\it l$^-$}}
\newcommand{\lp}{\mbox{\it l$^+$}}
\newcommand{\lpm}{\mbox{\it l$^{\pm}$}}
\newcommand{\lmp}{\mbox{\it l$^{\mp}$}}
\newcommand{\gel}{\mbox{$e$}}
\newcommand{\agel}{\mbox{$\overline{e}$}}
\newcommand{\elp}{\mbox{$e^+$}}
\newcommand{\elm}{\mbox{$e^-$}}
\newcommand{\elpm}{\mbox{$e^{\pm}$}}
\newcommand{\elmp}{\mbox{$e^{\mp}$}}
\newcommand{\gmu}{\mbox{$\mu$}}
\newcommand{\mum}{\mbox{$\mu^-$}}
\newcommand{\mup}{\mbox{$\mu^+$}}
\newcommand{\mupm}{\mbox{$\mu^{\pm}$}}
\newcommand{\mump}{\mbox{$\mu^{\mp}$}}
\newcommand{\gtau}{\mbox{$\tau$}}
\newcommand{\agtau}{\mbox{$\overline{\tau$}}}
\newcommand{\taum}{\mbox{$\tau^-$}}
\newcommand{\taup}{\mbox{$\tau^+$}}
\newcommand{\tam}{\mbox{$\tau^-$}}
\newcommand{\tap}{\mbox{$\tau^+$}}
\newcommand{\tapm}{\mbox{$\tau^{\pm}$}}
\newcommand{\tamp}{\mbox{$\tau^{\mp}$}}
\newcommand{\gnu}{\mbox{$\nu$}}
\newcommand{\agnu}{\mbox{$\overline{\nu}$}}
\newcommand{\nul}{\mbox{$\nu_l$}}
\newcommand{\anul}{\mbox{$\overline{\nu}_l$}}
\newcommand{\nue}{\mbox{$\nu_e$}}
\newcommand{\num}{\mbox{$\nu_{\mu}$}}
\newcommand{\nut}{\mbox{$\nu_{\tau}$}}
\newcommand{\anue}{\mbox{$\overline{\nu}_e$}}
\newcommand{\anum}{\mbox{$\overline{\nu}_{\mu}$}}
\newcommand{\anut}{\mbox{$\overline{\nu}_{\tau}$}}
%
\newcommand{\ggam}{\mbox{$\gamma$}}
\newcommand{\aggam}{\mbox{$\overline{\gamma}$}}
\newcommand{\aB}{\mbox{$b\overline{b}$}}
%
\newcommand{\sbab}{\mbox{$\sigma_{b\overline{b}}$}}
\newcommand{\sj}{\mbox{$\sigma_{\psi}$}}
\newcommand{\st}{\mbox{$\sigma_T$}}
%
\newcommand{\gB}{\mbox{\it B}}
\newcommand{\agB}{\mbox{$\overline{\it B}$}}
\newcommand{\gBz}{\mbox{$B^{\circ}$}}
\newcommand{\agBz}{\mbox{$\overline{B}^{\circ}$}}
\newcommand{\gBp}{\mbox{$B^+$}}
\newcommand{\gBm}{\mbox{$B^-$}}
\newcommand{\Bu}{\mbox{$B_u$}}
\newcommand{\Bupm}{\mbox{$B^{\pm}_u$}}
\newcommand{\Bup}{\mbox{$B^{+}_{u}$}}
\newcommand{\Bum}{\mbox{$B^{-}_{u}$}}
\newcommand{\Bd}{\mbox{$B_d$}}
\newcommand{\aBd}{\mbox{$\overline{B}_d$}}
\newcommand{\Bdz}{\mbox{$B^{\circ}_d$}}
\newcommand{\aBdz}{\mbox{$\overline{B}^{\circ}_d$}}
\newcommand{\Bs}{\mbox{$B_s$}}
\newcommand{\aBs}{\mbox{$\overline{B}_s$}}
\newcommand{\Bsz}{\mbox{$B^{\circ}_s$}}
\newcommand{\aBsz}{\mbox{$\overline{B}^{\circ}_s$}}
\newcommand{\Bc}{\mbox{$B^{\pm}_c$}}
%
\newcommand{\Piz}{\mbox{$\pi^{\circ}$}}
\newcommand{\gPi}{\mbox{$\pi$}}
\newcommand{\Pip}{\mbox{$\pi^+$}}
\newcommand{\Pim}{\mbox{$\pi^-$}}
\newcommand{\Pipm}{\mbox{$\pi^{\pm}$}}
\newcommand{\Pimp}{\mbox{$\pi^{\mp}$}}
%
\newcommand{\gK}{\mbox{$K$}}
\newcommand{\agK}{\mbox{$\overline{K}$}}
\newcommand{\Kp}{\mbox{$K^+$}}
\newcommand{\Km}{\mbox{$K^{-}$}}
\newcommand{\Kpm}{\mbox{$K^{\pm}$}}
\newcommand{\Kep}{\mbox{$K^{\ast{+}}$}}
\newcommand{\Kez}{\mbox{$K^{\ast{0}}$}}
\newcommand{\Kepm}{\mbox{$K^{\ast{\pm}}$}}
\newcommand{\Kem}{\mbox{$K^{\ast{-}}$}}
\newcommand{\Ke}{\mbox{$K^{\ast{0}}$}}
\newcommand{\Kz}{\mbox{$K^0$}}
\newcommand{\aKz}{\mbox{$\overline{K}^0$}}
\newcommand{\aKzr}{\mbox{$\overline{K}^0_r$}}
\newcommand{\aKzl}{\mbox{$\overline{K}^0_l$}}
\newcommand{\Ks}{\mbox{$K_{S}$}}
\newcommand{\Ksz}{\mbox{$K_{S}^0$}}
\newcommand{\Kl}{\mbox{$K_{L}$}}
\newcommand{\Klz}{\mbox{$K_{L}^0$}}
\newcommand{\Kone}{\mbox{$K_1$}}
\newcommand{\Ktwo}{\mbox{$K_2$}}
\newcommand{\epepp}{\mbox{$\epsilon^{\prime}/\epsilon$}}
%
\newcommand{\Jpsi}{\mbox{J/$\psi$}}
\newcommand{\Psip}{\mbox{$\psi^{\prime}$}}
\newcommand{\Jp}{\mbox{$\psi(2S)$}}
\newcommand{\Chione}{\mbox{$\chi_1$}}
\newcommand{\Chitwo}{\mbox{$\chi_2$}}
%
\newcommand{\Rhoz}{\mbox{$\rho^{\circ}$}}
\newcommand{\Rhop}{\mbox{$\rho^+$}}
\newcommand{\Rhom}{\mbox{$\rho^-$}}
\newcommand{\Rhopm}{\mbox{$\rho^{\pm}$}}
\newcommand{\Etaz}{\mbox{$\eta^{\circ}$}}
\newcommand{\Etap}{\mbox{$\eta\prime$}}
%
\newcommand{\gD}{\mbox{$D$}}
\newcommand{\agD}{\mbox{$\overline{D}$}}
\newcommand{\Dp}{\mbox{$D^+$}}
\newcommand{\Dm}{\mbox{$D^-$}}
\newcommand{\Du}{\mbox{$D_u$}}
\newcommand{\aDu}{\mbox{$\overline{D}_u$}}
\newcommand{\Dstar}{\mbox{$D^{\ast}$}}
\newcommand{\aDstar}{\mbox{$\overline{D}^{\ast}$}}
\newcommand{\Dz}{\mbox{$D^{\circ}$}}
\newcommand{\aDz}{\mbox{$\overline{D}^{\circ}$}}
\newcommand{\Dstarz}{\mbox{$D^{\ast\circ}$}}
\newcommand{\aDstarz}{\mbox{$\overline{D}^{\ast\circ}$}}
\newcommand{\Dstarp}{\mbox{$D^{\ast{+}}$}}
\newcommand{\Dsm}{\mbox{$D^{-}_{s}$}}
\newcommand{\Dsp}{\mbox{$D^{+}_{s}$}}
\newcommand{\Lc}{\mbox{$\Lambda^+_c$}}
%
\newcommand{\Tu}{\mbox{$T_u$}}
\newcommand{\aTu}{\mbox{$\overline{T}_u$}}
\newcommand{\Tc}{\mbox{$T_c$}}
\newcommand{\aTc}{\mbox{$\overline{T}_c$}}
%
\newcommand{\rgP}{\mbox{$p$}}
\newcommand{\ragP}{\mbox{$\overline{p}$}}
\newcommand{\gP}{\mbox{$p$}}
\newcommand{\agP}{\mbox{$\overline{p}$}}
\newcommand{\gPpm}{\mbox{$p^{\pm}$}}
\newcommand{\gPmp}{\mbox{$p^{\mp}$}}
\newcommand{\Ppm}{\mbox{$p^{\pm}$}}
\newcommand{\Pmp}{\mbox{$p^{\mp}$}}
\newcommand{\gN}{\mbox{\it n}}
\newcommand{\agN}{\mbox{$\overline{\it n}$}}
\newcommand{\gL}{\mbox{$\Lambda$}}
\newcommand{\agL}{\mbox{$\overline{\Lambda}$}}
\newcommand{\Lz}{\mbox{$\Lambda^0$}}
\newcommand{\aLz}{\mbox{$\overline{\Lambda}$$^0$}}
\newcommand{\gXi}{\mbox{$\Xi$}}
\newcommand{\agXi}{\mbox{$\overline{\Xi}$}}
\newcommand{\Xipm}{\mbox{$\Xi^{\pm}$}}
\newcommand{\Ximp}{\mbox{$\Xi^{\mp}$}}
\newcommand{\Xim}{\mbox{$\Xi^-$}}
\newcommand{\aXim}{\mbox{$\overline{\Xi}$$^+$}}
\newcommand{\Xiz}{\mbox{$\Xi^0$}}
\newcommand{\aXiz}{\mbox{$\overline{\Xi}$$^0$}}
\newcommand{\Xizs}{\mbox{$\Xi^{{\ast}0}(1530)$}}
\newcommand{\gSig}{\mbox{$\Sigma$}}
\newcommand{\Sigz}{\mbox{$\Sigma^0$}}
\newcommand{\aSigz}{\mbox{$\overline{\Sigma}$$^0$}}
\newcommand{\Sigp}{\mbox{$\Sigma^+$}}
\newcommand{\aSigp}{\mbox{$\overline{\Sigma^+}$}}
\newcommand{\Sigm}{\mbox{$\Sigma^-$}}
\newcommand{\Sigpm}{\mbox{$\Sigma^{\pm}$}}
\newcommand{\gOm}{\mbox{$\Omega$}}
\newcommand{\agOm}{\mbox{$\overline{\Omega}$}}
\newcommand{\Omm}{\mbox{$\Omega^-$}}
\newcommand{\Ompm}{\mbox{$\Omega^{\pm}$}}
\newcommand{\Ommp}{\mbox{$\Omega^{\mp}$}}
\newcommand{\aOmm}{\mbox{$\overline{\Omega}$$^+$}}
%
\newcommand{\lB}{\mbox{$\Lambda_b$}}
%
\newcommand{\Xic}{\mbox{${\Xi}^{\circ}_{c}$}}
\newcommand{\aXic}{\mbox{$\overline{\Xi}^{\circ}_{c}$}}
\newcommand{\btoba}{\makebox{$|\langle\abd|\bd\rangle|^2$}}
\newcommand{\batob}{\makebox{$|\langle\bd|\abd\rangle|^2$}}
\newcommand{\btob}{\makebox{$|\langle\bd|\bd\rangle|^2$}}
\newcommand{\batoba}{\makebox{$|\langle\abd|\abd\rangle|^2$}}
\newcommand{\Psipm}{\makebox{$\Psi_{\pm}$}}
\newcommand{\Psit}{\makebox{$\Psi(t)$}}
\newcommand{\Psiat}{\makebox{$\bar{\Psi}(t)$}}
\newcommand{\bzb}{\makebox{$|B^{\circ}\rangle$}}
\newcommand{\bzab}{\makebox{$|\bar{B}^{\circ}\rangle$}}
%
%
\newcommand{\pol}{\makebox{$\vec{P}$}}
\newcommand{\poll}{\makebox{$\vec{P}_{\Lambda}$}}
\newcommand{\polxi}{\makebox{$\vec{P}_{\Xi}$}}
\newcommand{\polp}{\makebox{$\vec{P}_p$}}
\newcommand{\pold}{\makebox{$\vec{P}_d$}}
%
%
\newcommand{\php}{\makebox{$\hat{p}_p$}}
\newcommand{\phd}{\makebox{$\hat{p}_d$}}
\newcommand{\ph}{\makebox{$\hat{p}$}}
\newcommand{\phl}{\makebox{$\hat{p}_{\Lambda}$}}
\newcommand{\phx}{\makebox{$\hat{p}_{\Xi}$}}
%
%
\newcommand{\galpha}{\mbox{$\alpha$}}
\newcommand{\agalpha}{\mbox{$\overline{\alpha}$}}
\newcommand{\gbeta}{\mbox{$\beta$}}
\newcommand{\agbeta}{\mbox{$\overline{\beta}$}}
\newcommand{\ggamma}{\mbox{$\gamma$}}
\newcommand{\aggamma}{\mbox{$\overline{\gamma}$}}
\newcommand{\gphi}{\mbox{$\phi$}}
\newcommand{\alp}{\makebox{$\alpha_p$}}
\newcommand{\bp}{\makebox{$\beta_p$}}
\newcommand{\gap}{\makebox{$\gamma_p$}}
\newcommand{\alal}{\makebox{$\alpha\alpha$}}
\newcommand{\alalbar}{\makebox{$\overline{\alpha}\overline{\alpha}$}}
\newcommand{\delalal}{\makebox{$\delta\alpha\alpha$}}
\newcommand{\delalalbar}{\makebox{$\delta\overline{\alpha}\overline{\alpha}$}}
\newcommand{\alxi}{\makebox{$\alpha_{\Xi}$}}
\newcommand{\aalxi}{\makebox{$\alpha_{\overline{\Xi}}$}}
\newcommand{\aalxibig}{\makebox{$\overline{\alpha}_{\Xi}$}}
\newcommand{\bxi}{\makebox{$\beta_{\Xi}$}}
\newcommand{\gaxi}{\makebox{$\gamma_{\Xi}$}}
\newcommand{\gamxi}{\makebox{$\gamma_{\Xi}$}}
\newcommand{\all}{\makebox{$\alpha_{\Lambda}$}}
\newcommand{\aall}{\makebox{$\alpha_{\overline{\Lambda}}$}}
\newcommand{\aallbig}{\makebox{$\overline{\alpha}_{\Lambda}$}}
\newcommand{\alom}{\makebox{$\alpha_{\Omega}$}}
\newcommand{\aalom}{\makebox{$\alpha_{\overline{\Omega}}$}}
\newcommand{\aalombig}{\makebox{$\overline{\alpha}_{\Omega}$}}
\newcommand{\betaom}{\makebox{$\beta_{\Omega}$}}
\newcommand{\delal}{\makebox{$\Delta\alpha_{\Lambda}$}}
\newcommand{\delaxi}{\makebox{$\Delta\alpha_{\Xi}$}}
\newcommand{\Axi}{\makebox{$A_{\Xi}$}}
\newcommand{\Aaxi}{\makebox{$A_{\overline{\Xi}}$}}
\newcommand{\Aratio}{\makebox{$\frac{\alpha + \overline{\alpha}}%
                                    {\alpha - \overline{\alpha}}$}}
\newcommand{\Aratiobig}{\makebox{$\D\frac{\alpha + \overline{\alpha}}%
                                         {\alpha - \overline{\alpha}}$}}
\newcommand{\Axiratio}{\makebox{$\frac{\alxi + \aalxi}{\alxi - \aalxi}$}}
\newcommand{\Axiratiobig}{\makebox{$\D\frac{\alxi + \aalxibig}{\alxi - \aalxibig}$}}
\newcommand{\Axiratiotxt}{\makebox{$(\alxi + \aalxi)/(\alxi - \aalxi)$}}
\newcommand{\Axiratiobigtxt}{\makebox{$(\alxi+\aalxibig)/(\alxi-\aalxibig)$}}
\newcommand{\bpxi}{\makebox{$B^{\prime}_{\Xi}$}}
\newcommand{\Al}{\makebox{$A_{\Lambda}$}}
\newcommand{\Aal}{\makebox{$A_{\overline{\Lambda}}$}}
\newcommand{\Alratio}{\makebox{$\frac{\all + \aall}{\all - \aall}$}}
\newcommand{\Alratiobig}{\makebox{$\D\frac{\all + \aalLbig}{\all - \aallbig}$}}
\newcommand{\Alratiotxt}{\makebox{$(\all + \aall)/(\all - \aall)$}}
\newcommand{\Alratiobigtxt}{\makebox{$(\all + \aallbig)/(\all - \aallbig)$}}
\newcommand{\Axil}{\makebox{${A}_{\Xi\Lambda}$}}
\newcommand{\delAxil}{\makebox{$\Delta{A}_{\Xi\Lambda}$}}
\newcommand{\deldelta}{\makebox{$\Delta\delta$}}
\newcommand{\Axilratio}{\makebox{$\frac{\alxi\all - \aalxi\aall}
                                       {\alxi\all + \aalxi\aall}$}}
\newcommand{\Axilratiobig}{\makebox{$\D\frac{\alxi\all - \aalxibig\aallbig}
                                            {\alxi\all + \aalxibig\aallbig}$}}
\newcommand{\Axilratiotxt}{\makebox{$(\alxi\all-\aalxi\aall)/
                                     (\alxi\all+\aalxi\aall)$}}
\newcommand{\Axilratiotxtbig}{\makebox{$(\alxi\all-\aalxibig\aallbig)/
                                     (\alxi\all+\aalxibig\aallbig)$}}
\newcommand{\Aom}{\makebox{$A_{\Omega}$}}
\newcommand{\aAom}{\makebox{$A_{\overline{\Omega}}$}}
\newcommand{\Aomratio}{\makebox{$\frac{\alom + \aalom}{\alom - \aalom}$}}
\newcommand{\Aomratiobig}{\makebox{$\D\frac{\alom + \aalombig}{\alom - \aalombig}$}}
\newcommand{\Aomratiotxt}{\makebox{$(\alom + \aalom)/(\alom - \aalom)$}}
\newcommand{\Aomratiotxtbig}{\makebox{$(\alom+\aalombig)/(\alom-\aalombig)$}}
\newcommand{\Aoml}{\makebox{${A}_{\Omega\Lambda}$}}
\newcommand{\Aomlratio}{\makebox{$\frac{\alom\all - \aalom\aall}
                                       {\alom\alL + \aalOm\aalL}$}}
\newcommand{\Aomlratiobig}{\makebox{$\D\frac{\alom\all - \aalombig\aallbig}
                                            {\alom\all + \aalombig\aallbig}$}}
\newcommand{\Aomlratiotxt}{\makebox{$(\alom\all-\aalom\aall)/
                                     (\alom\all+\aalom\aall)$}}
\newcommand{\Aomlratiotxtbig}{\makebox{$(\alom\all-\aalombig\aallbig)/
                                        (\alom\all+\aalombig\aallbig)$}}
\newcommand{\abar}{\makebox{$\overline{a}$}}
\newcommand{\leeyang}{\makebox{$\displaystyle
    \pold = \frac{(\alp + \polp\cdot\phd)\phd +
                     \bp(\polp{\times}\phd) +
                     \gap(\phd{\times}(\polp{\times}\phd))}
                     {(1{+}\alp\polp\cdot\phd)}$}}
\newcommand{\leeyangxi}{\makebox{$\displaystyle
    \polL = \frac{(\alxi + \polxi\cdot\phl)\phl +
                     \bxi(\polxi{\times}\phL) +
                     \gaxi(\phl{\times}(\polxi{\times}\phl))}
                     {(1{+}\alxi\polxi\cdot\phl)}$}}
\newcommand{\alphaly}{\makebox{$\displaystyle
    \alpha = \frac{2\mbox{Re}(S^{\ast}P)}{|S|^2 + |P|^2}$}}
\newcommand{\alphalyt}%
{\makebox{$\alpha = {2\mbox{Re}(S^{\ast}P)}/(|S|^2 + |P|^2)$}}
\newcommand{\betaly}{\makebox{$\displaystyle
    \beta  = \frac{2\mbox{Im}(S^{\ast}P)}{|S|^2 + |P|^2}$}}
\newcommand{\gammaly}{\makebox{$\displaystyle
    \gamma = \frac{|S|^2 - |P|^2}{|S|^2 + |P|^2}$}} 
\newcommand{\hypdk}{\makebox{$\displaystyle
    \frac{dP}{d\Omega} =
    \frac{1}{4\pi}(1 + \alpha\vec{P}_p{\cdot}\hat{p}_d)$}}
\newcommand{\hypdkp}{\makebox{$\displaystyle
    \frac{dP}{d\cos\theta} =
    \frac{1}{2}(1 + \alpha_pp_p\cos\theta$)}}
\newcommand{\hypdkn}{\makebox{$\displaystyle
    \frac{dN}{d\Omega} =
    \frac{N_0}{4\pi}(1 + \alpha\vec{P}_p{\cdot}\hat{p}_d)$}}
\newcommand{\hypdkxil}{\makebox{$\displaystyle
    \frac{dP}{d\cos\theta} =
    \frac{1}{2}(1 + \alpha_{\Xi}\alpha_{\Lambda}\cos\theta)$}}
\newcommand{\hypdkxiln}{\makebox{$\displaystyle
    \frac{dN}{d\cos\theta} =
    \frac{N_0}{2}(1 + \alpha_{\Xi}\alpha_{\Lambda}\cos\theta)$}}
%
\newcommand{\alphaom}{\makebox{$\alpha_{\Omega}$}}
\newcommand{\alphaaom}{\makebox{$\overline{\alpha}$$_{\Omega}$}}
\newcommand{\alphal}{\makebox{$\alpha_{\Lambda}$}}
\newcommand{\alphaal}{\makebox{$\overline{\alpha}$$_{\Lambda}$}}
\newcommand{\alphaxi}{\makebox{$\alpha_{\Xi}$}}
\newcommand{\alphaaxi}{\makebox{$\overline{\alpha}$$_{\Xi}$}}
%
%
\def\kpll{K^+ \rightarrow \pi^+ l^+ l^-}
\def\kmll{K^- \rightarrow \pi^- l^+ l^-}
\def\kpmumu{K^+ \rightarrow \pi^+ \mu^+ \mu^-}
\def\kmmumu{K^- \rightarrow \pi^- \mu^+ \mu^-}
\def\kpmmumu{K^{\pm} \rightarrow \pi^{\pm} \mu^{+}\mu^{-}}
\def\kpee{K^+ \rightarrow \pi^+ e^+ e^-}
\def\kp3pi{K^+ \rightarrow \pi^+ \pi^+ \pi^-}
\def\km3pi{K^- \rightarrow \pi^- \pi^- \pi^+}
\def\kpm3pi{K^{\pm} \rightarrow \pi^{\pm} \pi^+ \pi^-}
\def\kpi{K^{\pm}_{\pi 3}}
\def\kmu{K^{\pm}_{\pi \mu \mu}}
\def\rkppmm{\Gamma(K^+ \rightarrow \pi^+\mu^+\mu^-)}
\def\rkmpmm{\Gamma(K^- \rightarrow \pi^-\mu^+\mu^-)}
\def\rkpmpmm{\Gamma(K^{\pm} \rightarrow \pi^{\pm}\mu^+\mu^-)}
\def\rkpall{\Gamma(K^+ \rightarrow all)}
\def\rkmall{\Gamma(K^- \rightarrow all)}
\def\rkpmall{\Gamma(K^{\pm} \rightarrow all)}
\def\delkpm{\Delta(K^{\pm} \rightarrow \pi^{\pm}\mu^+\mu^-)}
\def\rkppmm{\Gamma(K^+_{\pi\mu\mu})}
\def\rkmpmm{\Gamma(K^-_{\pi\mu\mu})}
\def\rkpmpmm{\Gamma(K^{\pm}_{\pi\mu\mu})}
\def\delkpm{\Delta(K^{\pm}_{\pi\mu\mu})}
\def\kpmu{K^+_{\pi\mu\mu}}
\def\kmmu{K^-_{\pi\mu\mu}}
\def\kpmmu{K^{\pm}_{\pi\mu\mu}}
\def\kppi{K^+_{{\pi}3}}
\def\kmpi{K^-_{{\pi}3}}
\def\kpmpi{K^{\pm}_{{\pi}3}}
\def\nokpmu{N^{obs}_{K^+_{\pi\mu\mu}}}
\def\nokmmu{N^{obs}_{K^-_{\pi\mu\mu}}}
\def\nokpmmu{N^{obs}_{K^{\pm}_{\pi\mu\mu}}}
\def\nokppi{N^{obs}_{K^+_{\pi{3}}}}
\def\nokmpi{N^{obs}_{K^-_{\pi{3}}}}
\def\nokpmpi{N^{obs}_{K^{\pm}_{\pi{3}}}}
\newcommand{\missmassformula}{
      \[
         M^2_{\rm miss} = M^2_K(1 - \frac{p_{\pi}}{p_K}) +
                          m^2_{\pi}(1 - \frac{p_K}{p_{\pi}}) -
                          p_{\pi}p_K\theta^2
      \]
}
%
%
\newcommand{\mudif}{\makebox{$\mu^- \rightarrow e^-\nu_{\mu}\overline{\nu}_{e}$}}
\newcommand{\mudio}{\makebox{$\mu^- N_{\rm A,Z} \rightarrow e^-\nu_{\mu}\overline{\nu}_{e} N_{\rm A,Z}$}}
\newcommand{\muc}{\makebox{$\mu^- N_{\rm A,Z} \rightarrow \nu_{\mu} N_{\rm A,Z-1}$}}
\newcommand{\murc}{\makebox{$\mu^- N_{\rm A,Z} \rightarrow \nu_{\mu}\gamma N_{\rm A,Z-1}$}}
\newcommand{\mueg}{\makebox{$\mu \rightarrow e\gamma$}}
\newcommand{\mue}{\makebox{$\mu^-N \rightarrow e^-N$}}
\newcommand{\mueaz}{\makebox{$\mu^-N_{\rm A,Z} \rightarrow e^-N_{\rm A,Z}$}}
\newcommand{\mues}{\makebox{$\mu^- N_{\rm A,Z} \rightarrow e^- N_{\rm A,Z}$}}
\newcommand{\rmue}{\makebox{$R_{\mu e} = \frac{\Gamma (\gmu N \ra \gel N)}{\Gamma (\gmu N \ra \num N^{\ast})}$}}
\newcommand{\rmuetxt}{\makebox{$R_{\mu e} = 
   {\Gamma (\gmu N \ra \gel N)}/{\Gamma (\gmu N \ra \num N^{\ast})}$}}

%
\newcommand{\pienu}{\makebox{$\pi^- \rightarrow e^-\overline{\nu}_{e}$}}
\newcommand{\pirc}{\makebox{$\pi^- N_{\rm A,Z} \rightarrow \gamma N_{\rm A,Z-1}$}}



\begin{center}
{\LARGE \bf The \nova\ Power Distribution System} \\[0.2in]
\renewcommand{\thefootnote}{\fnsymbol{footnote}}
E.C. Dukes$^{c,}$\footnote[1]{Corresponding author.}, 
R. Ehrlich$^{c}$, 
S. Goadhouse$^{c}$, 
L. Mualem$^{a}$, 
A. Norman$^{b,}$\footnote[2]{Current address: Fermi National Accelerator Laboratory, Batavia, Illinois 60510, USA.}, 
and R. Tesarek$^{b}$ \\[0.1in]
{\footnotesize
$^{a}${\it California Institute of Technology, Pasadena, California 91125, USA} \\
$^{b}${\it Fermi National Accelerator Laboratory, Batavia, Illinois 60510, USA} \\
$^{d}${\it University of Virginia, Charlottesville, Virginia 22904, USA} \\[0.1in]
7 July 2018 \\[0.2in]
}
\end{center}

\begin{center}
\begin{minipage}{5.0in}
\hrule
\vspace*{0.1in}
\noindent {\bf Abstract} \\[0.05in]
\noindent We describe the power distribution systems and grounding schemes built for the near and far detectors
of the \nova\ long-baseline neutrino experiment.  They are used to power the 
avalanche photodiodes and their thermoelectric coolers, the front-end boards that read out, 
digitize and time stamp the signals from the avalanche photodiodes, and the data 
concentrator modules used to receive and format the data from the front-end boards 
before sending them to a farm of computers used to build the events.  The system powers 
344,064 readout channels in the far detector and 20,192 channels in the near detector. 
\\[0.1in]
\vspace*{0.1in}
\hrule
\end{minipage}
\end{center}
\renewcommand{\thefootnote}{\arabic{footnote}}

\section{Introduction}

\nova\ is a long-baseline neutrino experiment employing a 14\,kilotonne far detector (FD) near Ash
River, MN, USA and a smaller 290\,tonne near detector (ND) at Fermilab (Fig.~\ref{fig:nova_detectors}).  
Both detectors are centered at an angle of
14.6\,mrad from the axis of the Fermilab NuMI neutrino beam \cite{numi} and
are located 1\,km and 810\,km from the NuMI target.
The two detectors are essentially finely grained, low-Z, sampling calorimeters, 
with a high ratio of active to passive material.
They are functionally equivalent with the exception of the muon catcher
placed at the rear of the near detector.

The far detector is composed of 28 blocks, each consisting of 32 planes, for a total of 896 planes. 
Each plane contains 12 modules, the fundamental mechanical element of the detector.
Each module has 32 cells, the fundamental detector element, giving a total of 384 cells per plane.  
The orientation of the cells in each plane alternates between horizontal and vertical.
Each plane is offset from those of the same orientation adjacent to it by half a cell width.
The near detector has 6 blocks, each with 32 planes, giving a total of 192 planes.\footnote{The physical
and electrical definitions of a block differs for the near detector: we use the latter definition
throughout this paper.}
Each near detector plane consists of 3 modules, or 96 cells.
In addition, there is a muon catcher at the rear of the near detector that consists of
22 planes interspersed with 101.6-mm-thick (4\,in) iron plates.  
The muon catcher horizontal (vertical) oriented planes are 2 (3) modules wide for a total of 62 (96) cells per plane.
There are a total of 344,064 (20,192) cells in the far (near) detector.

Each cell is 39\,mm wide by 66\,mm deep (internal dimension) polyvinyl chloride (PVC) \cite{pvc} 
filled with liquid scintillator, pseudocumene being the main dopant \cite{scintillator}.  
Cell lengths are 15.5\,m in the far detector; 3.9\,m (2.6\,m in the vertical cells of muon catcher) in the near detector.
Each cell has a 0.7\,mm diameter wavelength-shifting fiber which is looped at the far end of the cell. 
The fiber transports light to an avalanche photodiode (APD) at the other end of the cell.
The 32 fibers of a module are read out by a single 32-pixel APD; both ends of a fiber are coupled to a single (rectangular) pixel.
They are cooled to $-15$\tc\ by thermoelectric coolers (TEC).
The signals from each APD pixel are continuously digitized by a custom front-end board (FEB) at 500\,ns intervals
for the far detector and 125\,ns for the near detector.
The FEBs also feed down-regulated high voltage to the APDs from a 425\,V input, 
and 24\,V to the thermoelectric cooler controller daughter board on the FEB that 
controls the temperature of the APD in a feedback loop.
The FEBs themselves are powered by 3.5\,V.  There are 10,752 (631) FEBs in the far (near) detector.
Zero-suppressed digitized signal snippets from the APDs are sent to data concentrator modules (DCM).
The DCMs, powered by 24\,V, accumulate 50\,$\mu$s data fragments into 5\,ms blocks from the 2048 cells served by 64 APDs/FEBs.
The DCMs then send their data via high-speed switches to a computer farm where
the full events are built and, if satisfying trigger criteria, sent to permanent storage \cite{daq}.

\begin{figure}[htb]
\centerline{\includegraphics[width=5.5in]{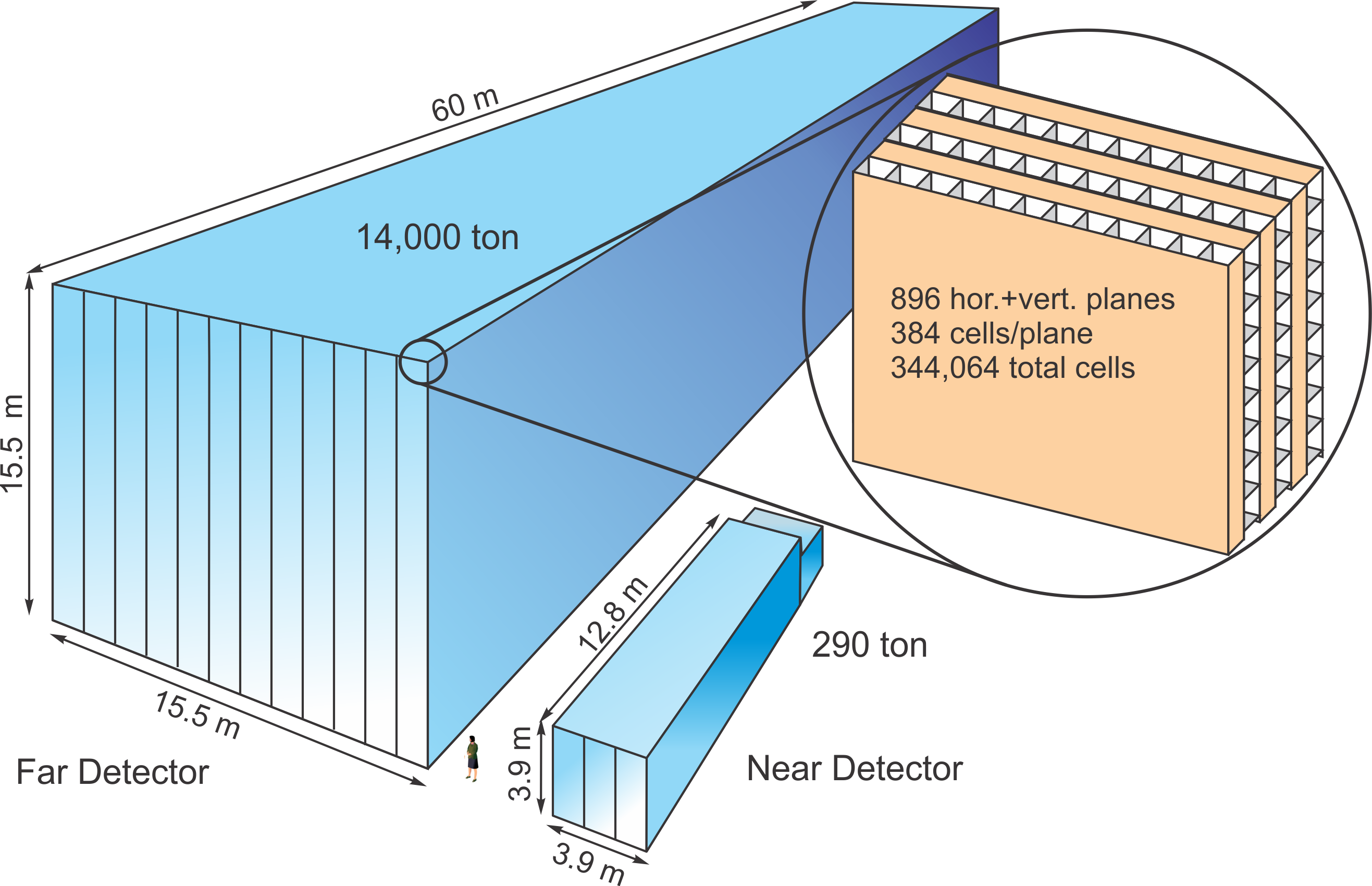}}
\caption[The \nova\ far and near detectors]%
{The \nova\ far and near detectors. The detectors have alternating vertical and horizontal PVC
cells filled with liquid scintillator.  Each plane is composed of twelve modules, each module
with 32 cells. 
}
\label{fig:nova_detectors}
\end{figure}

The \nova\ power distribution system supplies power to four different electronic
components used in the readout:  3.5\,V to the front-end boards, 425\,V to the avalanche photodiodes,
24\,V to the thermoelectric coolers, and 24\,V to the data concentrator modules.
It does {\em not} provide power to the timing distribution units used to provide an absolute time to the data,
which have an on-board AC-to-DC adapter that runs off of 120\,VAC.
The power distribution system consists of:  
(1) the power supplies and their relay racks, 
(2) the power distribution boxes that fan out the power to the FEBs, APDs, TECs, and DCMs, 
(3) the power cables and their cable trays, and
(4) the detector grounding system.
The numbers of each component are given in Table~\ref{tab:channel_counts}.
A detailed description of the system is found in Ref.~\cite{nova_pds}.
\begin{table}[htbp]
\footnotesize
\centering
\caption[\nova\ electronics channel counts]{\nova\ electronics channel counts.}
  \label{tab:channel_counts}
\vspace*{0.08in}
\begin{tabular}{lrr}
\toprule
  \multicolumn{1}{c}{Item} & Near Detector & Far Detector \\
\midrule
Channels (cells)                    & 20,192 & 344,064 \\
Front-end boards (FEB)              &    631 &  10,752 \\
Thermoelectric coolers (TEC)        &    631 &  10,752 \\
Avalanche photodiodes (APD)         &    631 &  10,752 \\
Data concentrator modules (DCM)     &     14 &     168 \\
Power distribution boxes            &     14 &     168 \\
Wiener MPOD high voltage mainframes &      1 &       2 \\
Wiener ISEG high voltage cards      &      1 &      11 \\
Wiener PL506 low voltage mainframes &      5 &      56 \\
Wiener PL506 2.0--5.8\,V pods       &     14 &     168 \\
Wiener PL506 12--30\,V  pods        &     14 &     168 \\
Relay racks                         &      3 &      16 \\
Total wall power                    &   6 kW &   71 kW \\
\bottomrule
\end{tabular}
\end{table}

The \nova\ power distribution system was largely designed by physicists, 
with help from several electrical engineers in key areas,
as electrical engineering resources were both scarce and costly. 
The system was designed to have very low noise levels, to be easily scalable, 
simple to operate, reliable, safe, and to have a lifetime in excess of ten years.
The large number of channels, extent of the detectors, and their remote locations
demanded that the power distribution system be remotely 
controllable.\footnote{The far and near detectors are operated remotely, usually without anyone on site.}
The very low light levels and low gain of the APDs demanded low-noise power supplies 
and that great care be taken in the design of the grounding system.
Estimates of the currents needed for the FEB, TEC, DCM, and APDs, based on early prototypes,
were 0.5\,A, 0.15\,A, 1.25\,A, and 40\,$\mu$A, respectively.
The system was designed to handle twice those currents in most cases, and indeed, the requirements increased
to the final values given in Table~\ref{tab:requirements}.
\begin{table}[htbp]
\footnotesize
\centering
\caption[Design power requirements for the detector electronics]{Design power requirements for the detector electronics.}
  \label{tab:requirements}
\vspace*{0.08in}
\begin{tabular}{ccc|ccc}
\toprule
  & & Max. & \multicolumn{3}{|c}{Power Distribution Box} \\
  Device & Voltage & Current & Channels & Total current & Power \\
\midrule
  FEB    &       3.5 V &  1.0 A      & 64     &   64 A   & 211 W \\
  TEC    &      24 V   &  0.30 A     & 64     &   20 A   & 480 W \\
  APD    &     425 V   &  40\,$\mu$A & 64     &  2.6 mA  &   2 W \\
  DCM    &      24 V   &  1.5 A      &  1     &  1.5 A   &  30 W \\
\bottomrule
\end{tabular}
\end{table}

\section{System Layout}

The layout for a far detector diblock is shown in Fig.~\ref{fig:layout_fd} 
and for the entire near detector in Fig.~\ref{fig:layout_nd}.
A system schematic of the power distribution system is shown in Fig.~\ref{fig:pds_layout} with
a more detailed electrical schematic given in Fig.~\ref{fig:pds_schematic}.
\begin{figure}[htbp]
\centerline{\includegraphics[width=5.5in]{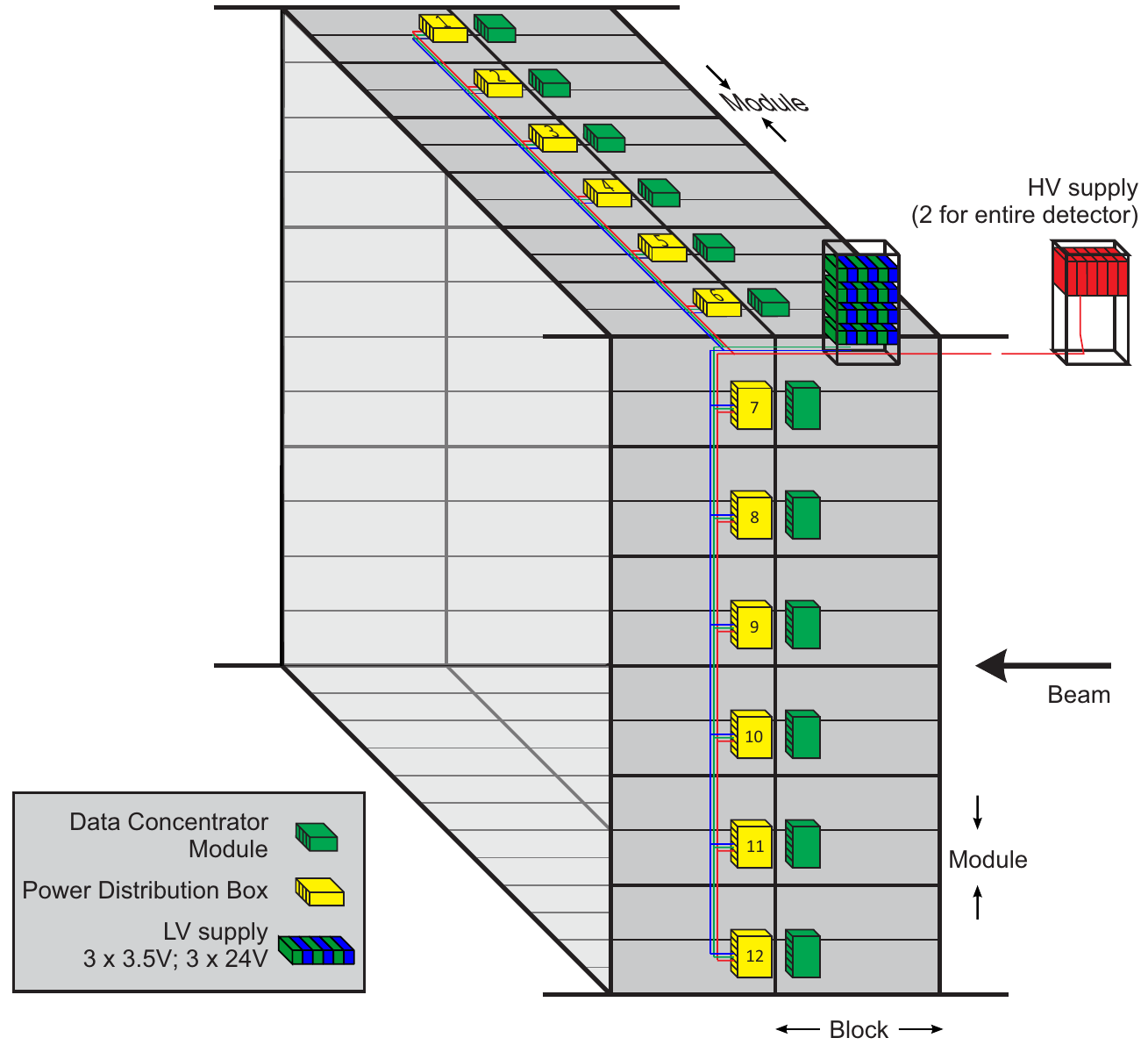}}
\caption[Layout of the power distribution system for a far detector diblock]%
{Layout of the power distribution system for one of the 14 far detector diblocks.  
Each of the 12 power distribution boxes powers all of the front-end boards (FEB) and data concentrator 
modules (DCM) in a two block deep by two module wide area.  The total for an entire diblock is
768 FEBs, 12 DCMs, and 24,576 channels (cells).  A relay rack on the upper catwalk hosts the low-voltage
power supplies needed for the diblock.  Two relay racks host the high-voltage power
supplies needed for all of the power distribution boxes for the entire detector.
}
\label{fig:layout_fd}
\end{figure}
\begin{figure}[htbp]
\centerline{\includegraphics[width=5.5in]{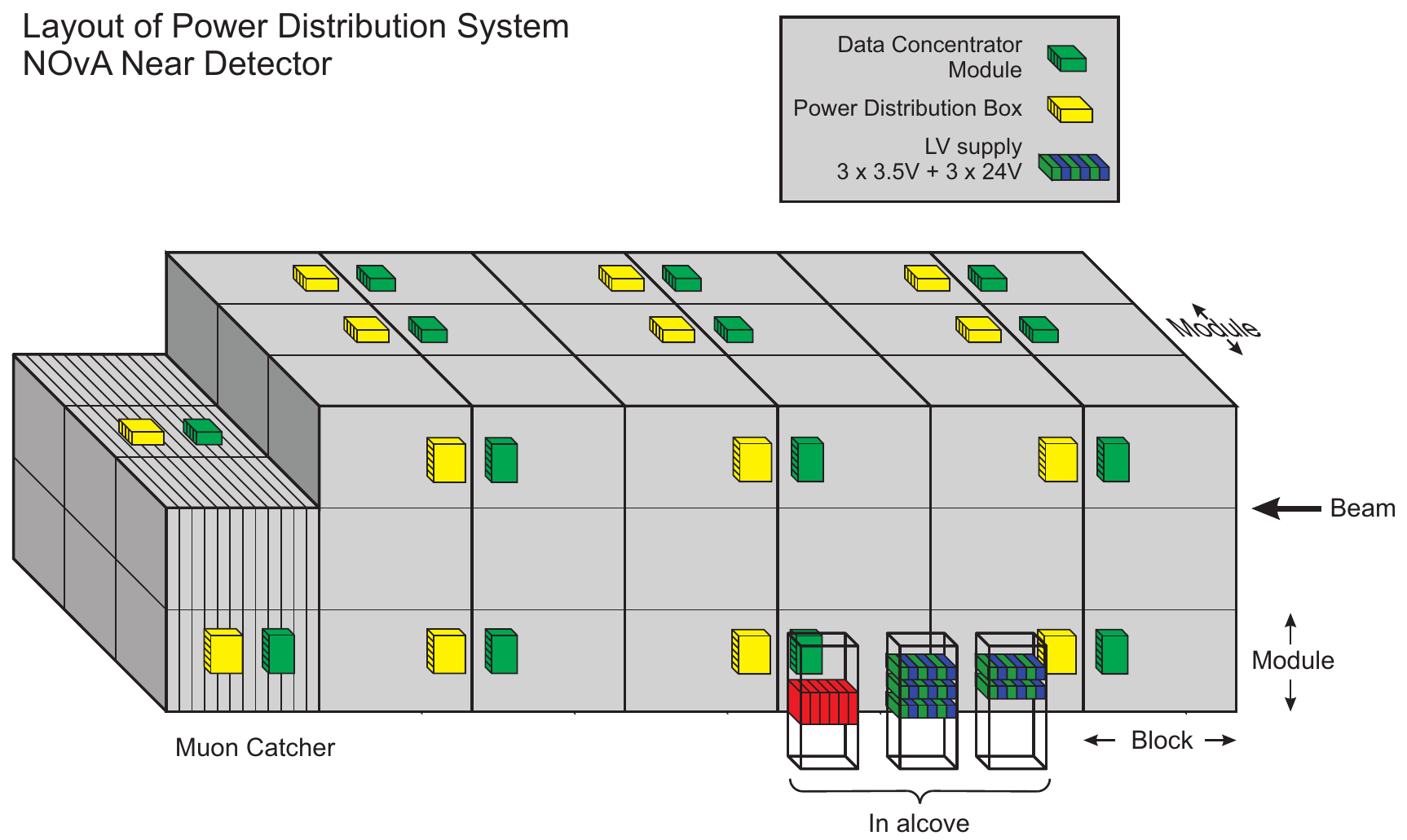}}
\caption[Layout of the power distribution system for the near detector]%
{Layout of the power distribution system for the near detector.
Each top/beam-left (bottom/beam-right) power distribution box feeds power to front-end boards in a two block by two (one) module wide portion of the detector.
Two power distribution boxes feed the muon catcher.
}
\label{fig:layout_nd}
\end{figure}

\begin{figure}[htb]
\centerline{\includegraphics[width=5.5in]{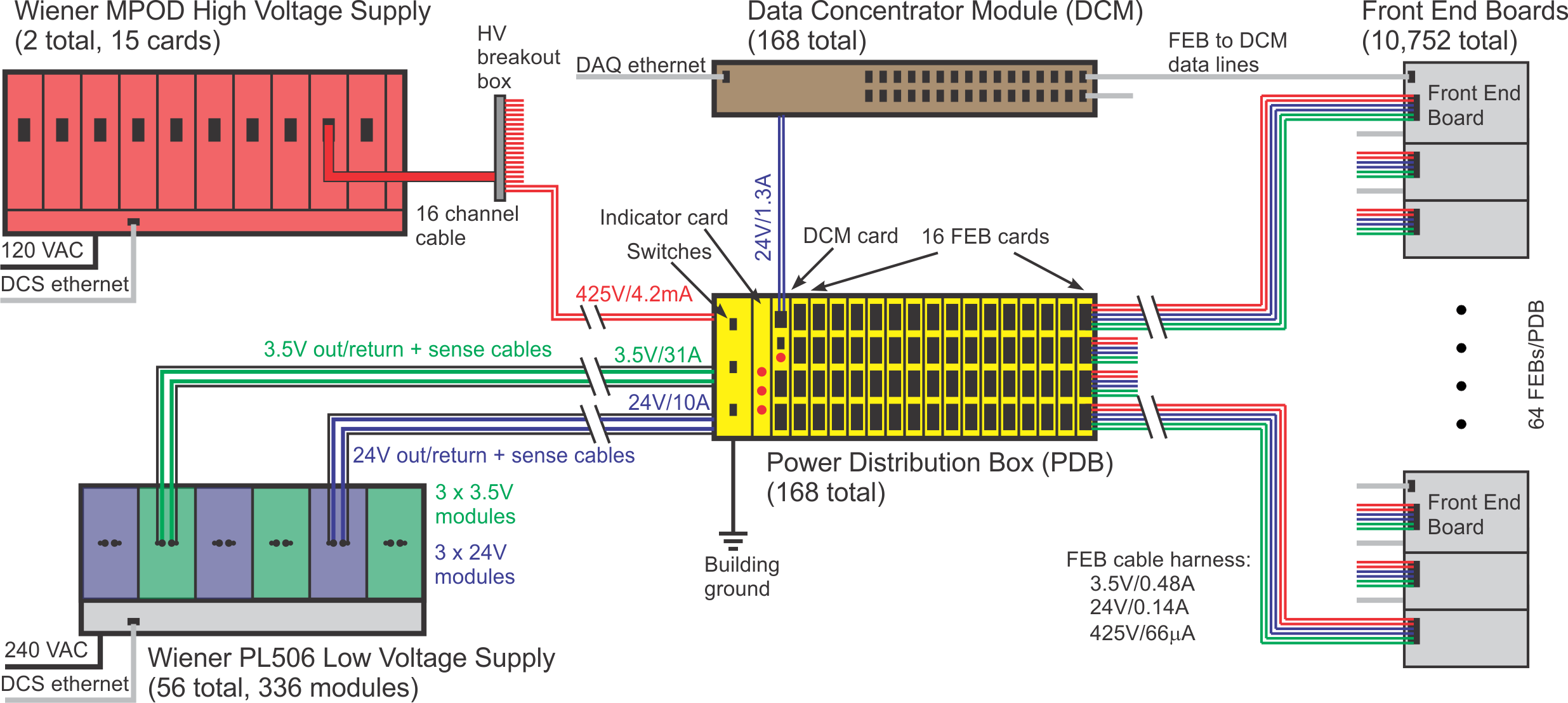}}
\caption[Layout of the power distribution system]%
{Layout of the power distribution system serviced by a single power distribution box.  
Each power distribution box feeds 3.5\,V, 24\,V, and 425\,V via 6-conductor cables 
to 64 front-end boards, and 24\,V via a 2-conductor cable to the nearby data concentrator module.  
The 425\,V to the power distribution boxes is provided by a Wiener MPOD mainframe.  The 3.5\,V and 24\,V 
power are provided by a Wiener PL506 power supply.  The channel counts shown are for the
far detector only.}
\label{fig:pds_layout}
\end{figure}

The repeatable unit served by the power distribution system is two blocks, called a diblock,
composed of 64 detector planes.
All power to the front-end electronics is supplied by low voltage and high voltage
power supplies in relay racks on the top catwalk of the far detector hall and in an alcove hewn into the rock
for the near detector.
The power is distributed by power distribution boxes which are placed on the detector.
Each far detector power distribution box feeds (via 16 four-channel cards) a 32 plane by two module wide segment of the detector:
64 FEBs, or 2048 APDs (see Fig.~\ref{fig:layout_fd}).
Each near detector power distribution box feeds a 32 plane by either a one or two module wide segment
of the detector: 64 or 32 FEBs (see Fig.~\ref{fig:layout_nd}).
The near detector muon catcher has a single power distribution box feeding the 33 FEBs that serve the vertical oriented modules
and another power distribution box that serves the 22 horizontal FEBs.
A total of 64 single 6-conductor, 18\,AWG cables from each power distribution box to 64 front-end boards carry the high voltage needed by the APDs,
the 24\,V needed by the TECs, and the 3.5\,V needed by the FEBs, as well as their return
currents.  All the electronic components and the power supplies float: ground reference is
at the power distribution boxes.\footnote{The high-voltage supply chassis for safety reasons are connected via a 1\,k$\Omega$ resistor
to the power distribution box ground through the high-voltage cable shield, as shown in Fig.~\ref{fig:pds_schematic}.} 
Note that having the ground reference at the front-end boards would have required equal-length PDB-to-FEB cables as well as
ground cables to all 10,752 front-end boards.

There is an associated data concentrator module (DCM) for each power distribution box,
whose power is also supplied through the power distribution box. 

A far detector diblock requires 12 power distribution boxes, each feeding power to 64 FEBs of the 32
vertical or horizontal planes they serve.
The low-power supplies are mounted on relay racks on the upper
catwalk:  one supply serves three power distribution boxes for a total of four for each diblock.  
Two high-voltage mainframes are needed for the entire detector: they are situated at the upper catwalk midpoint.  
Since each high-voltage channel supplies power to one power distribution box, 12 high voltage channels are needed to supply
the 12 power distribution boxes in a diblock.

A drawing of part of the far detector vertical modules serviced by a single power distribution box is shown in Fig.~\ref{fig:layout_top}.
The power distribution boxes and data concentrator modules were required to have a low profile (limited to 3U or 133.35\,mm) 
in order to allow the rolling access bridge used to install and service the electronics mounted on 
top of the detector to be as close as possible to the detector.
The power distribution boxes and data concentrator modules that serve the vertical modules are mounted on a custom-built tray \cite{apm} whose
feet rest on the horizontal modules (see Fig.~\ref{fig:pds_photo_top}).
The power distribution boxes (and data concentrator modules) serving the horizontal modules are mounted sideways on the catwalk side
of the detector using a commercial framing system (see Fig.~\ref{fig:pds_photo_side}).

\begin{sidewaysfigure}[htbp]
\centerline{\includegraphics[width=8.0in]{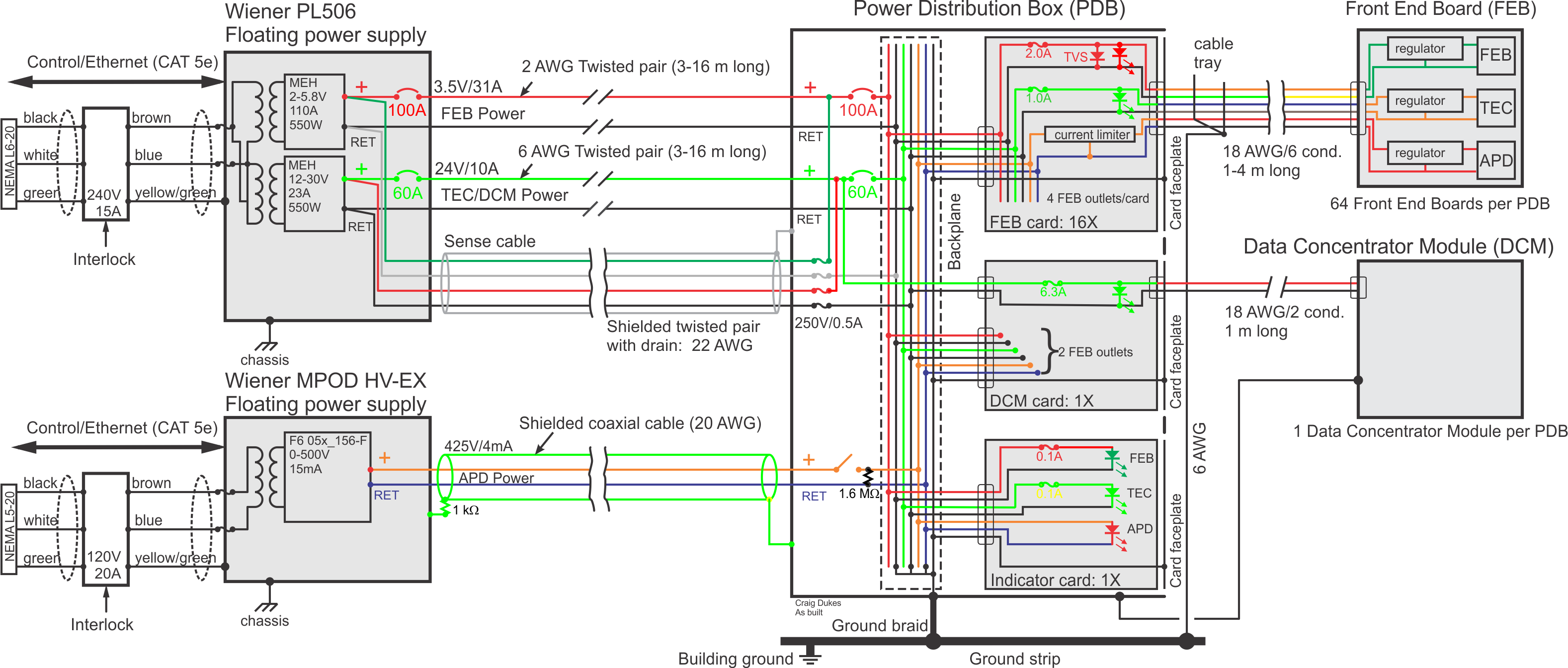}}
\caption[Electrical schematic of the power distribution system]%
{Electrical schematic of the power distribution system for a single power distribution box.
The low voltage and high voltage  power supplies float; ground reference is at the power distribution box.
Note that the high-voltage breakout box is not shown.}
\label{fig:pds_schematic}
\end{sidewaysfigure}

The layout for the power distribution system for the near detector is shown in Fig.~\ref{fig:layout_nd}.
The power distribution boxes are on the top and on one side of the detector.  
A total of 10 power distribution boxes, 4 low-voltage supplies, and one high-voltage mainframe with one high voltage card 
serve the entire near detector.  

\begin{figure}[htbp]
\centerline{\includegraphics[width=5.5in]{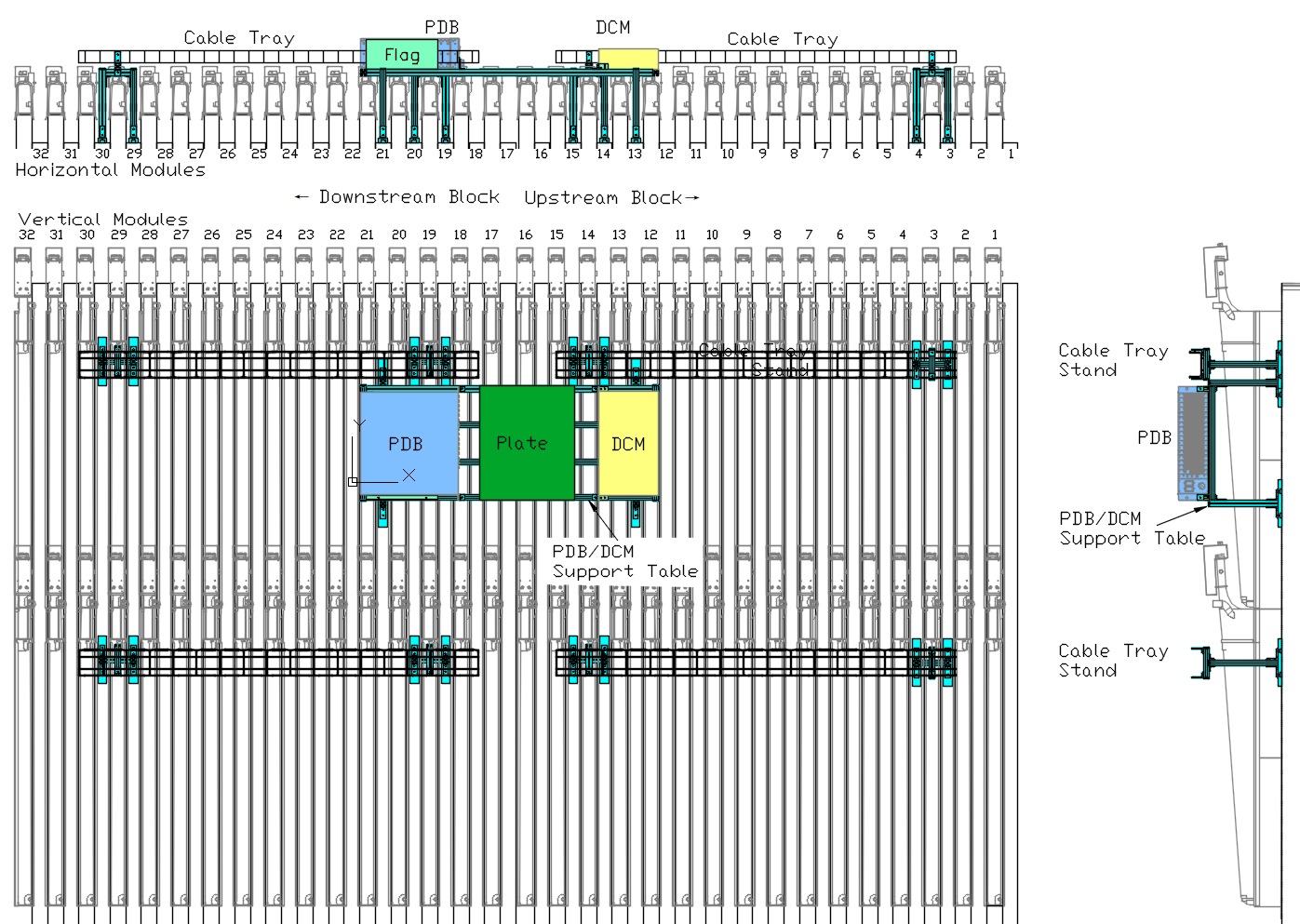}}
\caption[Front view of the top west corner of a far detector diblock]%
{Front, side, and top views of a far detector top two-module wide sector served by a single power distribution box.
The power distribution box, data concentrator module, support table, cable trays and their supports are shown;
cables are not.
}
\label{fig:layout_top}
\end{figure}

\begin{figure}[htbp]
\centerline{\includegraphics[width=5.5in]{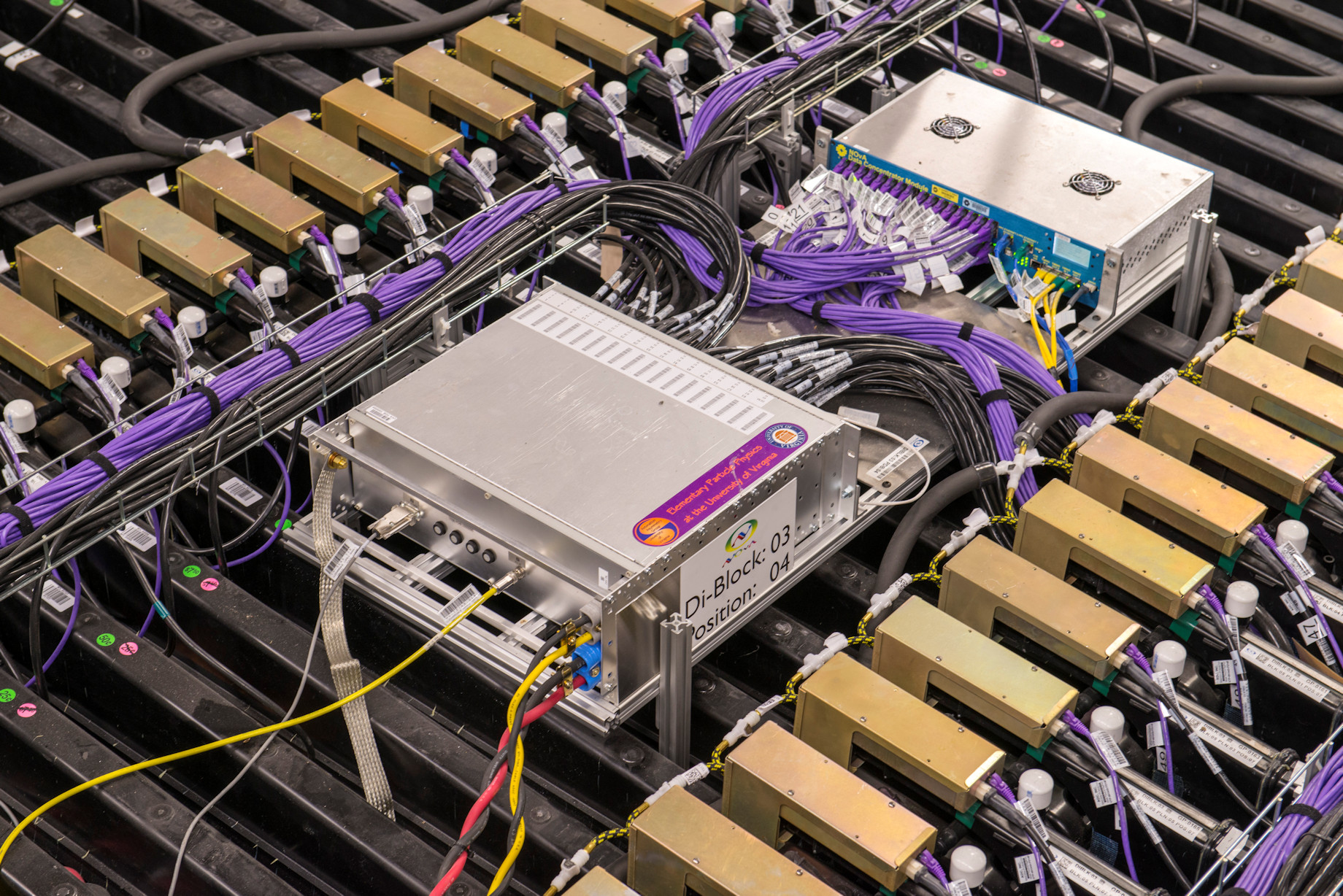}}
\caption[Photograph of a power distribution box and data concentrator module on top of the far detector]%
{Photograph of a power distribution box (foreground) and data concentrator module (background) 
on top of the far detector.  Black 6-conductor power cables run from the power 
distribution box to the front-end boards (copper).  
Violet CAT5e cables run from the data concentrator module to the font-end boards.  
The ground braid, sense cable (gray), high voltage cable (yellow),
3.5\,V cable, and 24\,V cable can be seen connected to the back of the power distribution box.
The front-end boards and APDs are inside the copper enclosures.
}
\label{fig:pds_photo_top}
\end{figure}

\begin{figure}[htbp]
\centerline{\includegraphics[width=5.5in]{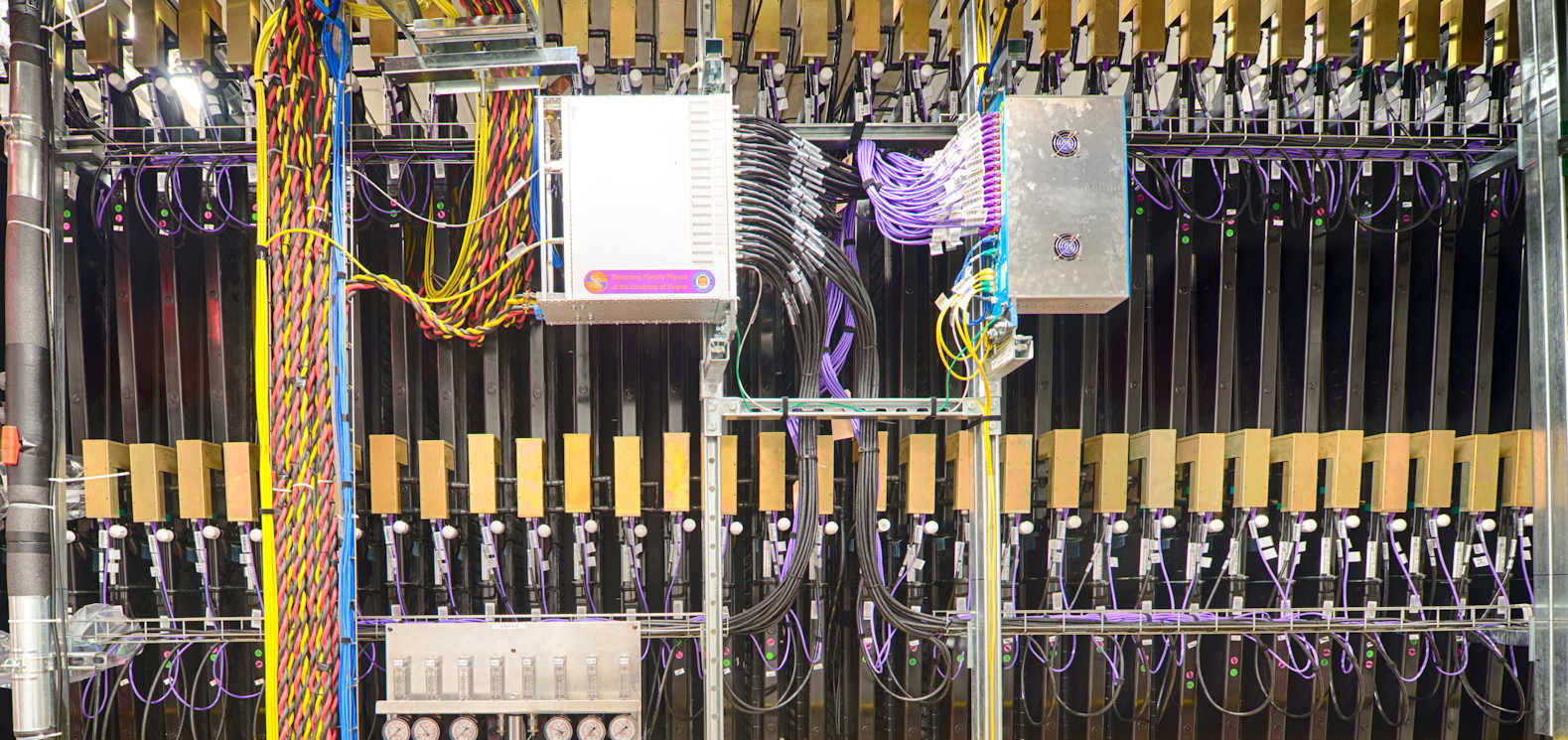}}
\caption[Photograph of a power distribution box and data concentrator module on the side of the far detector]%
{Photograph of a power distribution box and data concentrator module 
mounted on the side of the far detector.  
The power cables in the vertical cable tray at left come down from the power supplies located
in relay racks situated on the top catwalk.
}
\label{fig:pds_photo_side}
\end{figure}

\section{Power Supplies}

\subsection{High Voltage Power Supplies}

The high voltage to the power distribution boxes is provided by two Wiener \cite{wiener} MPOD HV-EX high voltage power supplies
outfitted with 15 ISEG EHS F6 05xi\_156-F floating 16-channel cards.
(See Table~\ref{tab:power_par} for the power supply parameters.)  
Each of the 16 channels  of
the ISEG card is individually programmable from 0--500\,V and provides a maximum of 15\,mA of current.
Input power to the supplies is single-phase, 120\,V, 20\,A.
A custom-made cable harness \cite{hvcable} takes the 16 high voltage channels from each ISEG
card to a nearby breakout box \cite{apm} (see Fig.~\ref{fig:hv_breakout_box}).
The breakout box has a jumper that if interrupted cuts the high-voltage power off at the supply.
Shielded triaxial RG-58A/U cables with triaxial 3-lug BNC connectors go from the high-voltage
breakout box to the power distribution boxes.
Each Wiener high voltage channel feeds a single power distribution box or up to 64 avalanche photodiodes.

\begin{figure}[htbp]
\centerline{\includegraphics[width=4.0in]{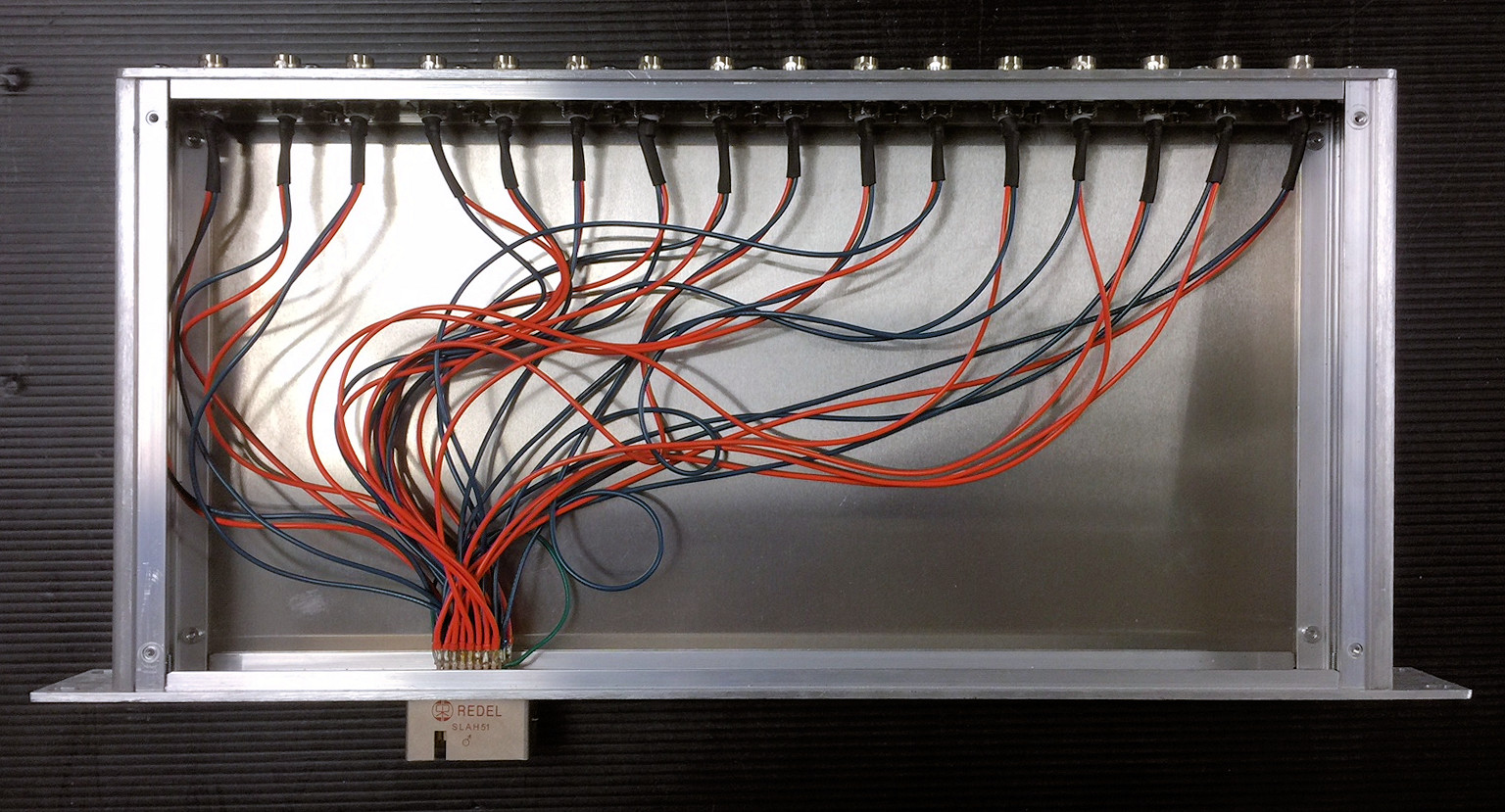}}
\caption[High voltage breakout box]
{Photograph of a high-voltage breakout box with the top cover off.
The triaxial connectors at top are mounted on an insulated circuit board and connected through 1\,k$\Omega$ 
resistors to ground, a safety requirement in case the cables were disconnected at the power distributions boxes
or perhaps accidentally cut.  The REDEL connector can be seen at bottom.
}
\label{fig:hv_breakout_box}
\end{figure}

\subsection{Low Voltage Power Supplies}

The required power for a far detector diblock, including cable losses, is given in Fig.~\ref{fig:pds_power}.
The low voltages (3.5\,V and 24\,V) needed by the FEBs, DCMs, and TECs, are provided by
the Wiener PL506 LX Power Supply System \cite{wiener}).  
This power supply was chosen for its low noise, high current capability, and high power density.
No specific noise requirements were given by the designers of the \nova\ front-end electronics;
however, the supplies were tested to insure that they added nothing to the noise budget.
The high power density is particularly important as the relay racks into which the power supplies are mounted
are full, and adding additional relay racks in the limited catwalk space was not possible.
Each Wiener PL506 chassis has six floating individually
programmable power supply pods: three 2.0--5.8\,V pods, each rated to 115\,A and 550\,W, and
three 12--30\,V pods, each rated to 23\,A and 550\,W.   (Note:  the standard PL506 LX low voltage
module has a range of 2--7\,V.  Because the front-end boards have a voltage regulator
that is rated only to 6.0\,V, the pods procured from Wiener were modified to
only go to 5.8\,V.)  
The power supplies have a remote sense feature which was used as the voltage drops between the power supplies
and the power distribution boxes are large and vary.
The remote sense feature can be run in fast, moderate, and slow modes: the moderate setting is used due to 
the lengths of the cables.\footnote{Note that when the remote sense setting was inadvertently set to a different mode the power supply tripped.}
Input power to the supplies is single-phase, 240\,V, 15\,A.
The 240\,V input was chosen early in the system design in order to allow the 3.5\,V and 24\,V pods to be run at their maximum power.

Each PL506 crate feeds power to three power distribution boxes:  the 3.5\,V via 2\,AWG cables and the 24\,V via 6\,AWG cables.  
The cables are rated to safely carry the maximum current of the Wiener supplies without the need for fuses;
nevertheless they were fused using a special box affixed
to the rear of the power supply mainframe \cite{apm} (see Fig.~\ref{fig:lv_breaker_box}).
The remote sensing feature allows the voltages at the power distribution boxes to be set to their desired values.  
This feature is needed because the different cable lengths from the low-voltage power supplies to 
the power distribution boxes produce non-negligible differences in the voltage drops.  
The longest (shortest) cable runs are 21\,m (6\,m), corresponding to one-way 
voltage drops of 0.32\,V (0.07\,V) and 0.23\,V (0.05\,V), respectively for the 3.5\,V and 24\,V lines.  
Voltage drops on the 18\,AWG cables running from the power distribution boxes to the FEBs are smaller, 
resulting in a voltage range at the FEBs between 3.45--3.40\,V, 
well within the allowed range of the FEB voltage regulator.
For each diblock the input wall power is 5084\,W of which 3866\,W (76\%) is delived to the
front-end boards and data concentrator modules, 187\,W (4\%) is lost in cables, and 12\,W (0.2\%) lost in the power distribution boxes.
The APDs consume no appreciable power.
A total of 71\,kW of power is used by the entire far detector system.
\begin{figure}[htbp]
\centerline{\includegraphics[width=4.0in]{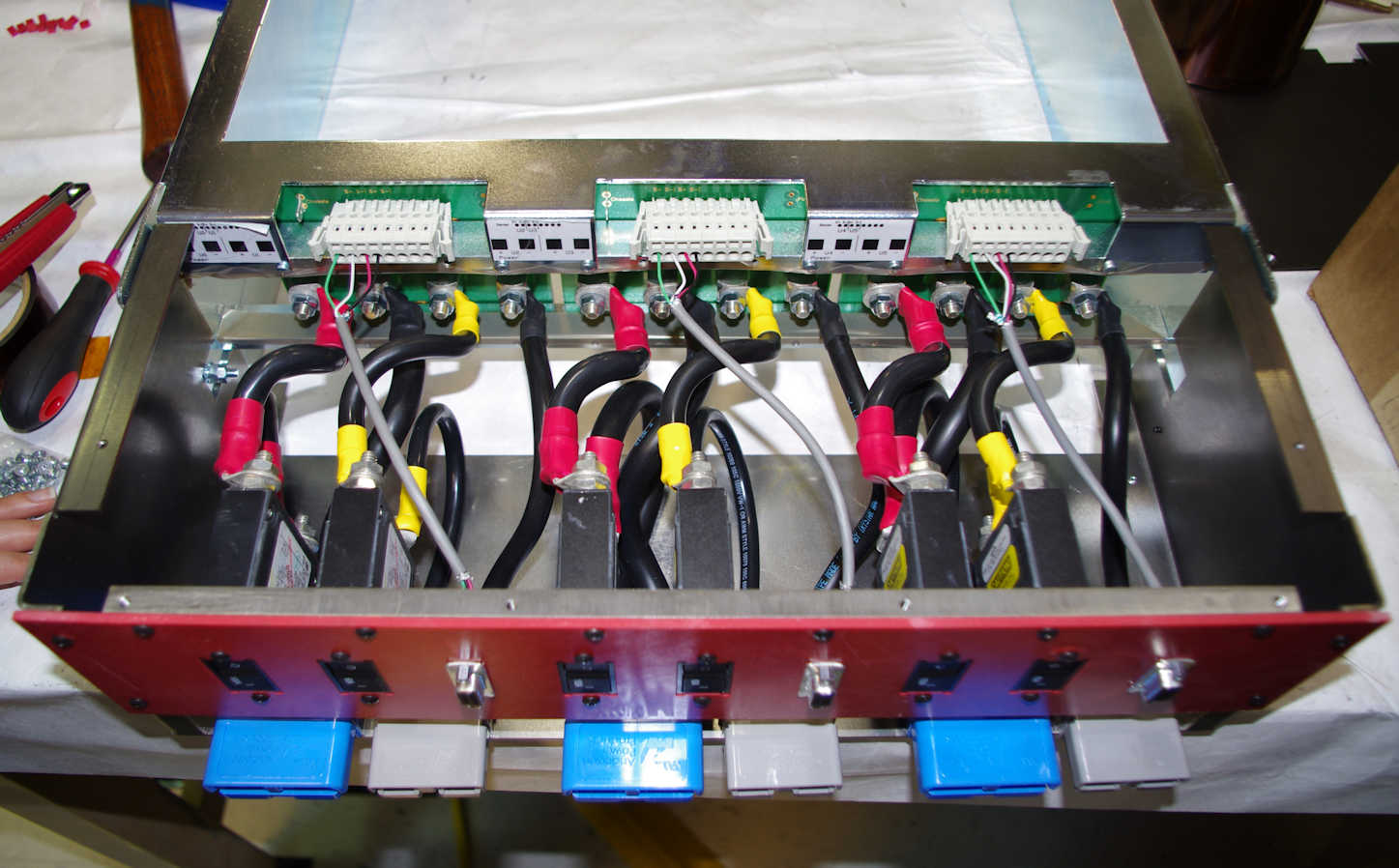}}
\caption[Low voltage breaker box]
{Photograph of the back of a low-voltage power supply showing the attached breaker box with the cover removed.
The 3.5\,V (24\,V) connectors can be seen in blue (gray) at the back.  Just inside are the 6 black breakers.
The sense cables are gray, the power cables are black, and the 3.5\,V (24\,V) cable lugs are red (yellow).
}
\label{fig:lv_breaker_box}
\end{figure}

\begin{table}[htbp]
\footnotesize
\centering
\begin{minipage}{5.5in}
\caption{Power supplies parameters.} 
\label{tab:power_par}
\vspace*{0.08in}
\begin{tabular}{rcc}
\toprule
  \multicolumn{1}{c}{Item}  &  Low Voltage                  & High Voltage \\
\midrule
  Manufacturer                &  Wiener                       &  Wiener  \\
  Mainframe type              &  PL506                        &  MPOD HV-EX  \\
  Size                        &  3U                           &  8U  \\
  Weight                      &  27.3 kg                      &  24 kg \\
  Power                       &  3840 W (240 VAC, 16\,A)      &  1200 W \\
                              &  550 W per pod                &         \\
  Input Power                 &  92 -- 265 VAC, 16\,A         &  95 -- 220 VAC, 50/60 Hz, 16\,A \\
  Power factor                &  0.997                        &  \\
  Efficiency                  &  77\% (\@4.2\,V)              &  88\% (\@24.8\,V) \\
  Pods (cards) per mainframe  &  6                            &  10  \\
  Pod (card) type             &  MEH                          &  ISEG EHS F6 05x\_156-F \\
  Channels per pod (card)     &  1                            &  16 \\
  Voltage range               &  2 -- 5.8\,V\footnote{Modified from standard 2 -- 7\,V.}
                                                              &  0 -- 500\,V \\
                              &  12 -- 30\,V                  &  \\
  Maximum current             &  115\,A (3.5\,V) 23\,A (24\,V)    &  15mA  \\
  Voltage resolution          &                               &  100\,mV \\
  Ripple                      &  $<$ 3\,mV pp                 &  $<$ 30\,mV pp \\
  Regulation                  &  $<$ 25\,mV @ $\pm$100\% load &   \\
  Remote sense                &  Fast:  $<$ 1\,m run           &  NA \\
                              &  Moderate: $>$ 1\,m  $<$ 30\,m run &  NA \\
                              &  Slow:   $>$ 30\,m run         &  NA \\
  Float range                 &  $\pm$10V                      & 200\,V  \\
\bottomrule
\end{tabular}
\end{minipage}
\end{table}

Both the low- and high-voltage Wiener power supplies are remotely controllable with Ethernet interfaces.
They have programmable trip levels.

\begin{sidewaysfigure}[htbp]
\centerline{\includegraphics[width=7.0in]{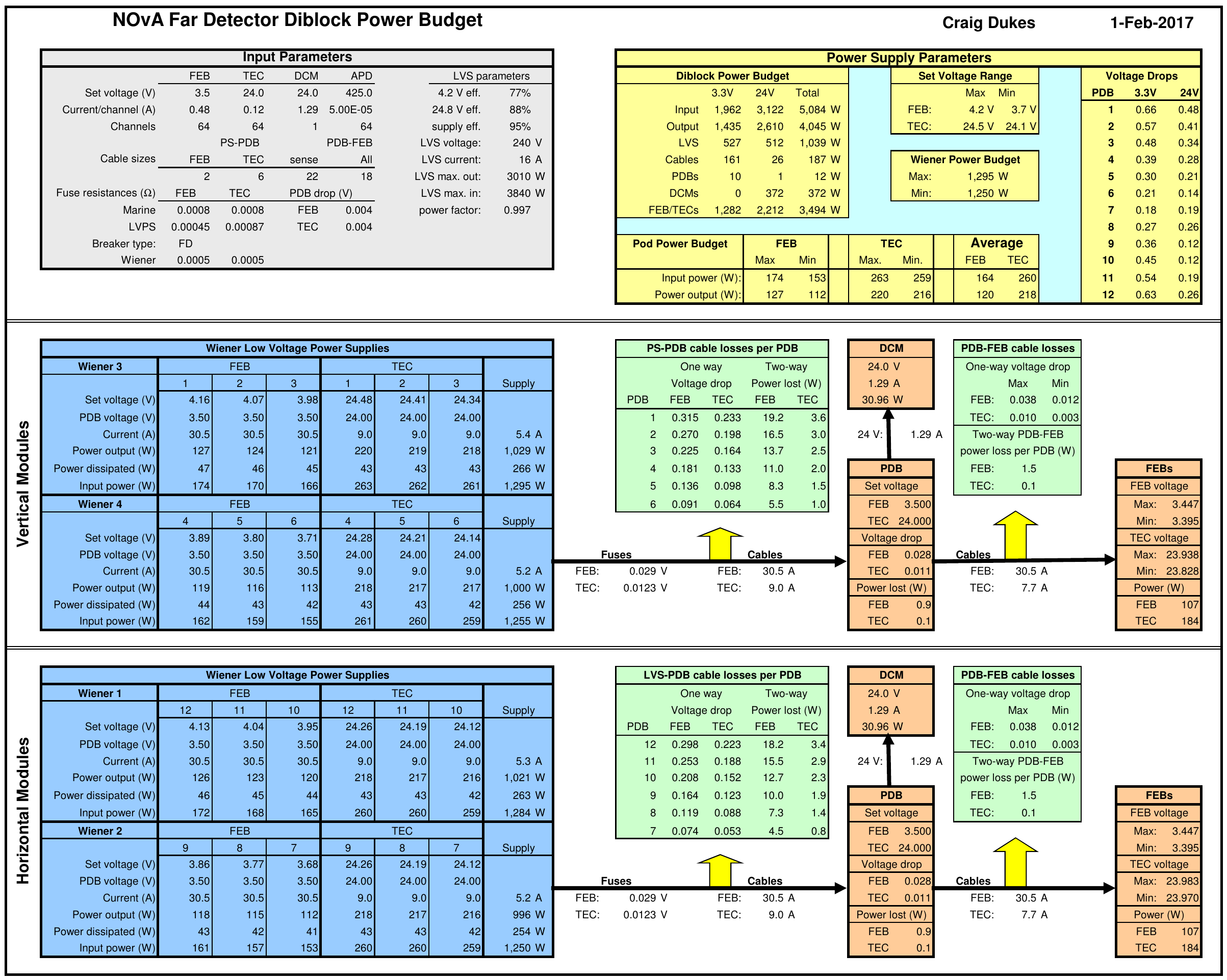}}
\caption[Power budget of the power distribution system]%
{Power budget for a far detector diblock.  
Included are the power supply losses, cable losses, and losses in the power distribution boxes.  
Two Wiener low-voltage supply mainframes serve the 6 power distribution boxes of the vertical modules, 
and two serve the 6 power distribution boxes of the horizontal modules.
}
\label{fig:pds_power}
\end{sidewaysfigure}

\subsection{Power Supply Testing}

The high- and low-voltage power supplies underwent extensive testing at the University of Virginia upon reception from the vendor
to insure they met requirements.
A separate test jig was used for each type of supply.  A technician and two undergraduate students performed the tests,
documented in Ref.~\cite{lv-ps-tests} and Ref.~\cite{hv-ps-tests}.

\subsubsection{Low Voltage Power Supply Testing}

The tests included verifying the proper operation of:
(1) the SNMP communication,
(2) the configuration settings,
(3) the voltage stability during a long burn-in period,
(4) the voltage accuracy and precision,
(5) temperature monitoring,
(6) the 5.8\,V limit for the 3.5\,V channels,
and
(7) the remote sense operation.
Ripple was also measured and was required to be less than 10\,mV (25\,mV) for the 3.3\,V (24\,V) channels.
The low-voltage power supply tests were run using a National Instruments PXIe-1062Q chassis running LabVIEW 2009 on Windows XP \cite{ni}.
Data were recorded and stored.
Each of the 6 pods in a Wiener PL506 low-voltage mainframe was connected to a BK Precision 8510 600\,W programmable
DC electronic load \cite{bkprecision} via 15.2\,m-long (50\,ft) 2 and 6 AWG cables.
It was found that the source and return cables had to be wound around each other at one twist per 0.3\,m (1\,ft)
to eliminate a loud audible sound caused the cable inductance effect on the remote sense system.
For the voltage stability tests the channels were burned in for three hours at 90\% capacity and the 
output voltage drift was recorded.

Of the 70 power supply mainframes and 420 pods that were tested, 8 pods failed the due to voltages deviating by more than 25\,mV from
their set point values.  Two of these were new pods, and six were older pods used for prototype tests.
After recalibration all supplies passed all tests.
In general the Wiener power supplies were found to work extremely well.

\subsubsection{High Voltage Power Supply Testing}

The high-voltage power supply tests included tests of the: voltage stability, voltage accuracy and precision, and
general functionality.
The tests were run using a National Instruments PXIe-1062Q chassis running LabVIEW 2009 on Windows XP.
Data were recorded and stored.
GwInstek PEL-2041 high-voltage programmable loads \cite{gwinstek}, as well as a custom-made fixed load for
burn-in tests, were used.

Of the 15 high-voltage cards tested, 13 passed all tests.
One had two channels that fell outside of the set value by more than the 145\,mV specified by the manufacturer.
The other had a channel that would not ramp up to the proper value.
The cards were replaced by the manufacturer.
The five mainframes all passed the tests as well as the 13 custom-made, high-voltage breakout boxes.

\section{Power Distribution Boxes}

\subsection{Overview}

There are a total of 168 (14) far (near) detector power distribution boxes.
Figures \ref{fig:pdb_isometric_front} and \ref{fig:pdb_isometric_back} show front and rear views of a power distribution box.
To fit within the tight space constraints of both detectors the power distribution boxes 
employ a standard DIN 3U subrack \cite{schroff}.
A custom backplane feeds 3.5\,V, 24\,V, and 425\,V to 18 cards:  
16 FEB cards that feed 3.5\,V, 24\,V, and 425\,V power each to four front-end boards, one DCM card that 
feeds 24\,V power to a data concentrator module, and one Indicator card that has LED lamps that 
indicate of the on/off status of the power distribution box.
Schematics of circuit boards used in the power distribution box can be found in Ref.~\cite{pdb_schematics}.
Figure~\ref{fig:pdb_layout_channel_feb} and Fig.~\ref{fig:pdb_layout_channel_dcm} show the front-panel layout
of the FEB and DCM cards;
Fig.~\ref{fig:card_feb}, Fig.~\ref{fig:card_dcm}, and Fig.~\ref{fig:card_indicator} show photographs of the front and back
of the FEB, DCM and Indicator cards described below.
\begin{figure}[htbp]
\centerline{\includegraphics[width=4.5in]{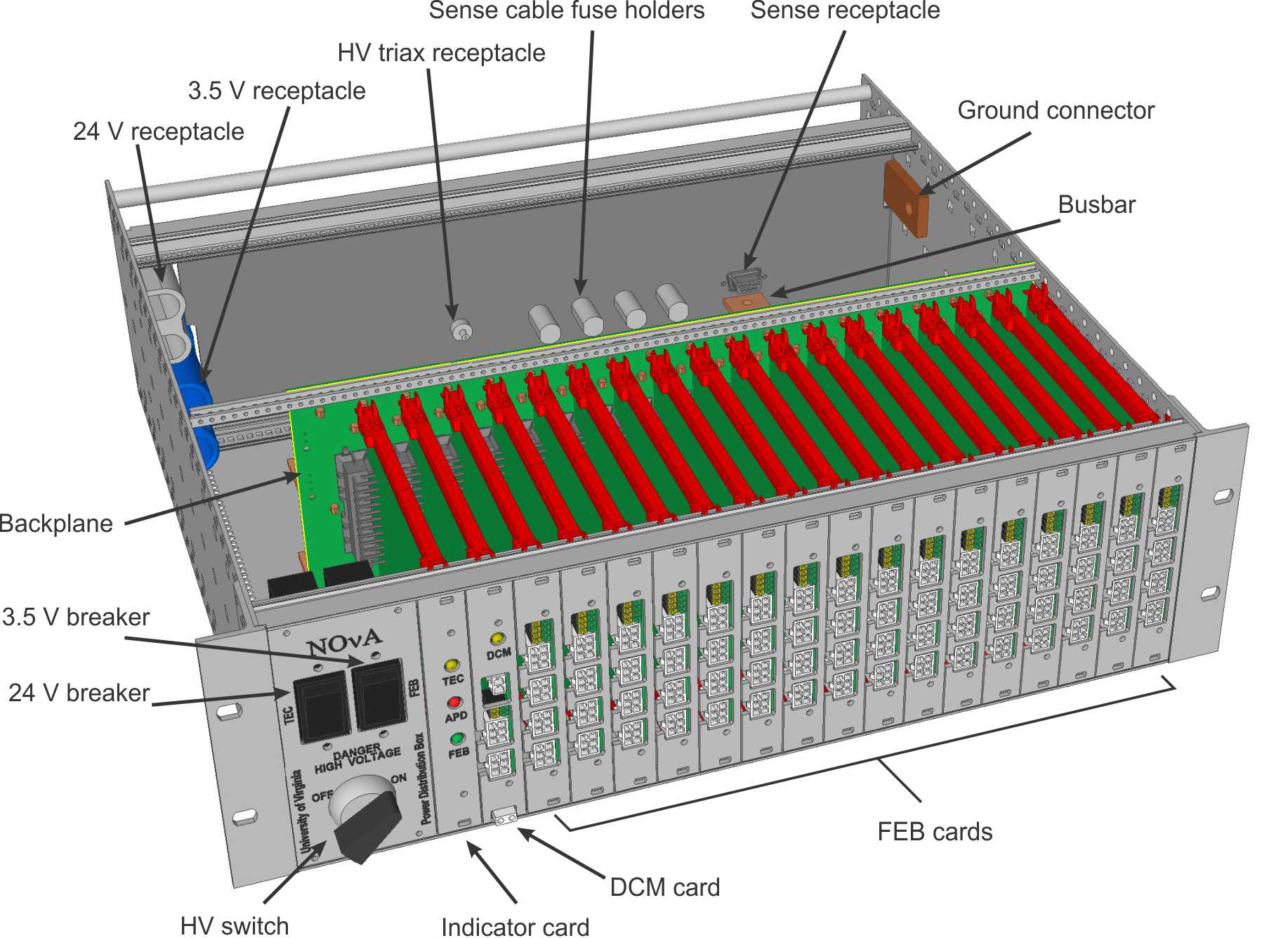}}
\caption[CAD drawing of the front of the power distribution box]%
{CAD drawing of the front of the power distribution box.  The top is removed to show the inside.
Internal cables are not shown.
}
\label{fig:pdb_isometric_front}
\end{figure}
\begin{figure}[htbp]
\centerline{\includegraphics[width=4.5in]{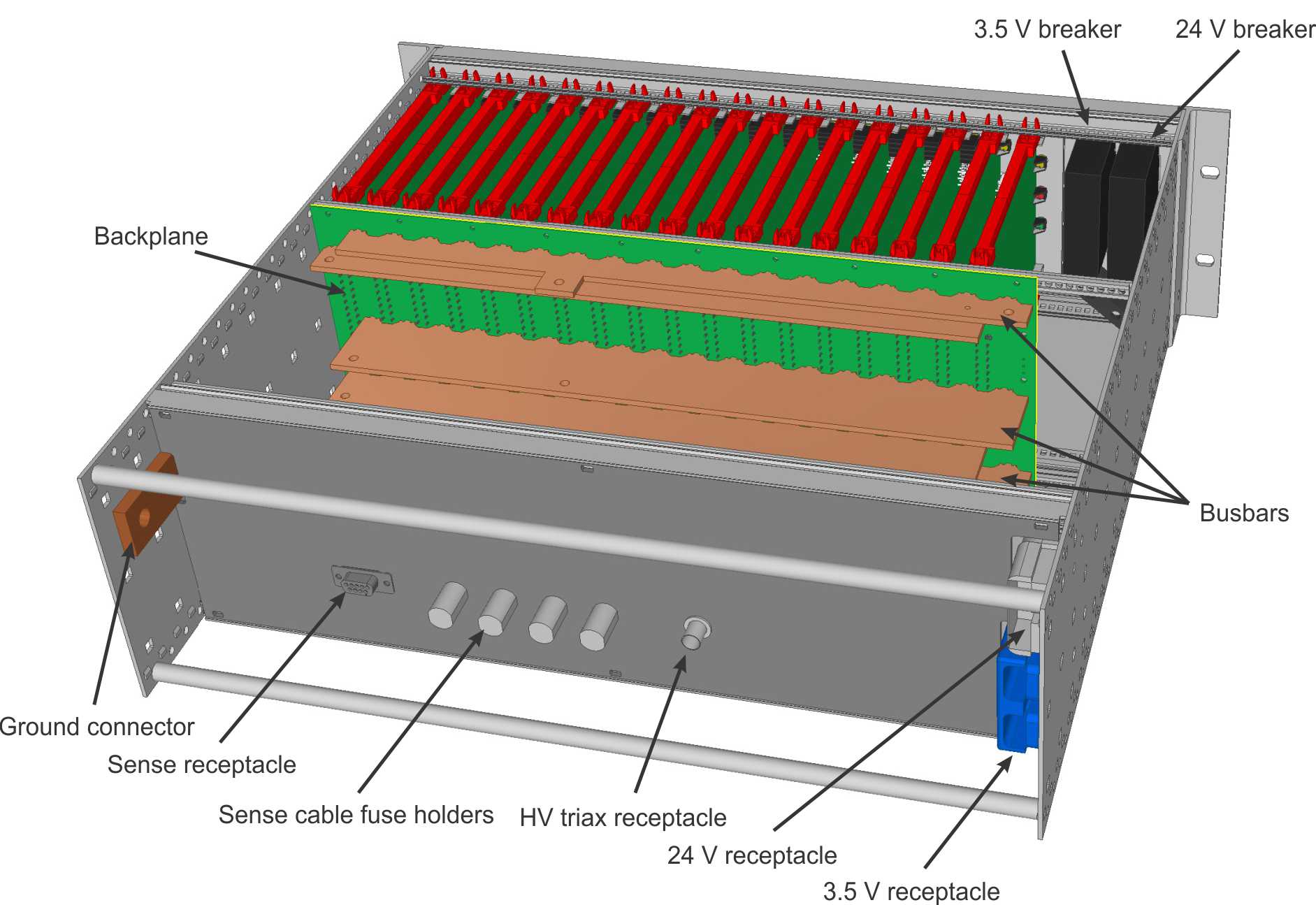}} 
\caption[CAD drawing of the back of the power distribution box]%
{CAD drawing of the back of the power distribution box.The top is removed to show the inside.
Internal cables are not shown.
}
\label{fig:pdb_isometric_back}
\end{figure}

The low-voltage power is fed to the power distribution boxes through quick connect receptacles at 
the back \cite{anderson}.  
Internal cables go to front-panel circuit breakers rated at 100\,A and 60\,A respectively for the 3.5\,V and
24\,V lines \cite{marine}.  
These breakers serve as fuses, but can also be used to locally power off the power distribution box, although the
operating procedure is to always power off the crate at the supply.
The high-voltage power goes to a front-panel switch rather than a breaker.

The power distribution box was designed to provide the currents given in Table~\ref{tab:pdb_currents},
all of which are roughly a factor of two greater than the original design currents.
The individual cards, described below, are fused at 1/0.75 = 1.33 of the maximum rating.
The card trace widths are set to handle currents 1.4 times the fused value.
The power distribution boxes and their cards were fabricated by a local electronics manufacturing firm near the 
University of Virginia \cite{www}.  They were tested at the University of Virginia using a custom
test stand described below.

\begin{table}[htbp]
\footnotesize
\centering
\caption[PDB currents]{Power distribution box current design capacity.}
  \label{tab:pdb_currents}
\vspace*{0.08in}
\begin{tabular}{lccc}
  \toprule
\multicolumn{1}{c}{Item} &   24\,V & 3.5\,V  & 425\,V \\
\midrule
Backplane                &   72\,A & 132\,A  & NA      \\
FEB card 1 channel       &  1.0\,A & 2.0\,A  & 300\,$\mu$A \\
FEB card 4 channels      &  4.0\,A & 8.0\,A  & 1.2\,mA      \\
DCM card output to DCM   &  6.3\,A &  NA   & NA      \\
\bottomrule
\end{tabular}
\end{table}

Providing higher voltage DC power to the power distribution boxes and then stepping it down using DC-to-DC converters
for the 3.5\,V and 24\,V lines was considered.  
Although such a system would have been in principle less expensive, 
sparse engineering design resources would have had to been used,
and the cost, including design, would have been essentially the same.
Note that there are no space constraints limiting the size of the power cables to the power distribution boxes.
Producing clean, noise-free power was also important; DC-to-DC converters would have added noise to the system.

The power distribution boxes were designed from the outset to be simple distribution boxes 
that fan out power to the FEB boards, with no means by which the outputs could be switched
remotely on or off.
To do so the appropriate cable to the front-end board must be unplugged.  
Operationally, there has been no need for such a feature.

\subsection{Backplane}

The backplane feeds power to the all of the cards.
The 3.5\,V, 24\,V, and ground feeds are copper busbars \cite{storm} that are soldered to the backplane 
circuit board. 
They range in size from 3.2$\times$28.6\,mm$^2$ (1/8\inch$\times$1.125\inch) for the 3.5\,V source and return;
3.2$\times$15.9\,mm$^2$ (1/8\inch$\times$0.625\inch) for the 24\,V source and return; and
3.2$\times$41.3\,mm$^2$ (1/8\inch$\times$1.625\inch)  for the ground.
The 3.5\,V return and 24\,V return lines are connected to the ground busbar, although if it
were decided to float the power distribution box rather than the power supplies, those connections can be easily disconnected.  
To reduce the possibility of a short, the return busbars are on the outside.  
The busbar cross sectional area was set to limit the current to be below the 1.55\,A/mm$^2$ (1000\,A/in$^2$)
requirement given in Ref.~\cite{fnal_design_standards}.
Note that the busbar current capacities exceed the maximum output capacities of the power supplies.

\begin{figure}[htbp]
\centering
\begin{minipage}[t]{2.5in}
\centerline{\includegraphics[type=png,ext=.png,read=.png,width=2.4in]{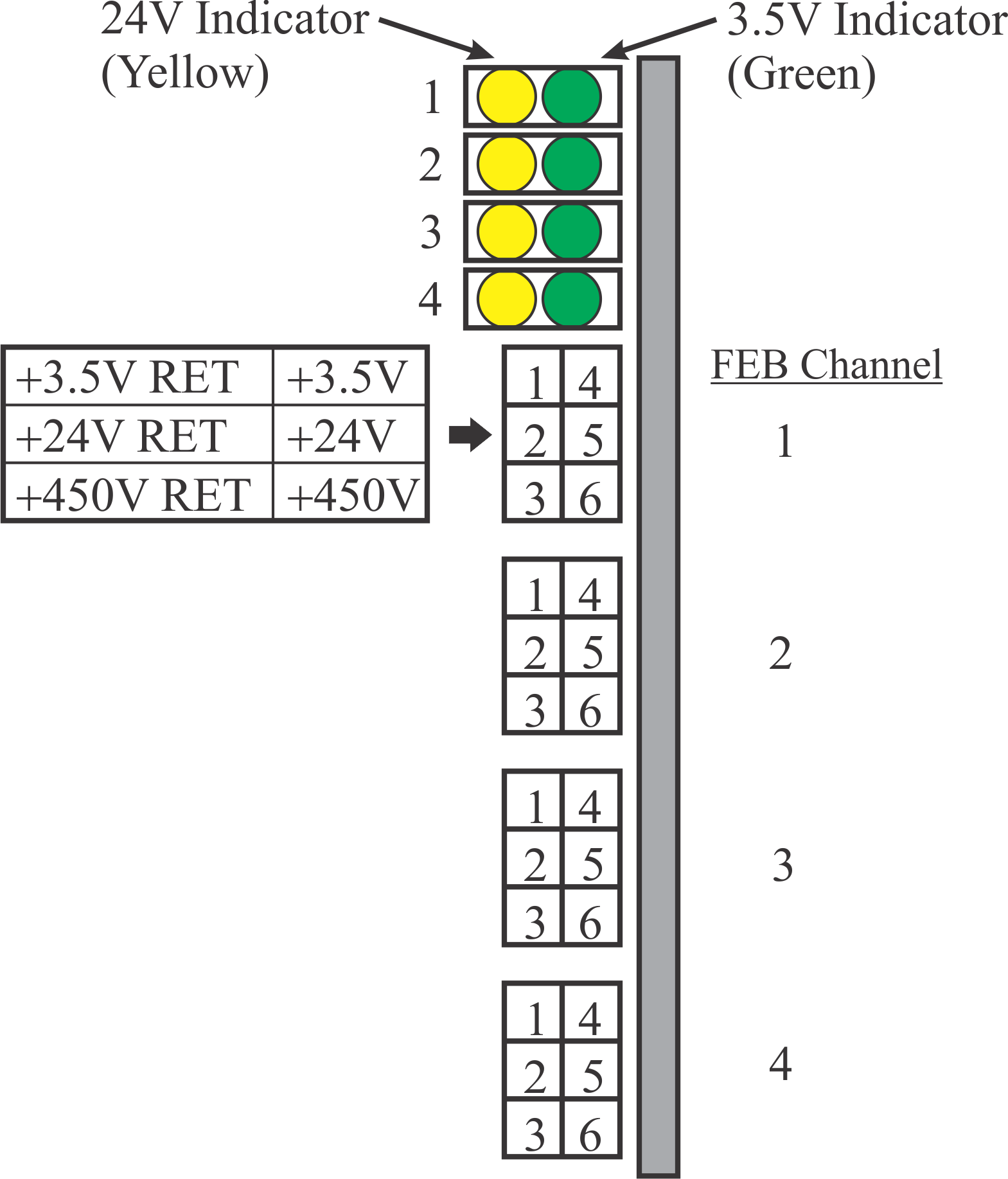}}
\caption[Front panel layout of power distribution box FEB card]%
{Front panel layout of power distribution box FEB card.  It feeds 3.5\,V, 24\,V and 425\,V to four FEBs.}
\label{fig:pdb_layout_channel_feb}
\end{minipage}
\hspace*{0.1in}%
\begin{minipage}[t]{2.5in}
\centerline{\includegraphics[type=png,ext=.png,read=.png,width=2.4in]{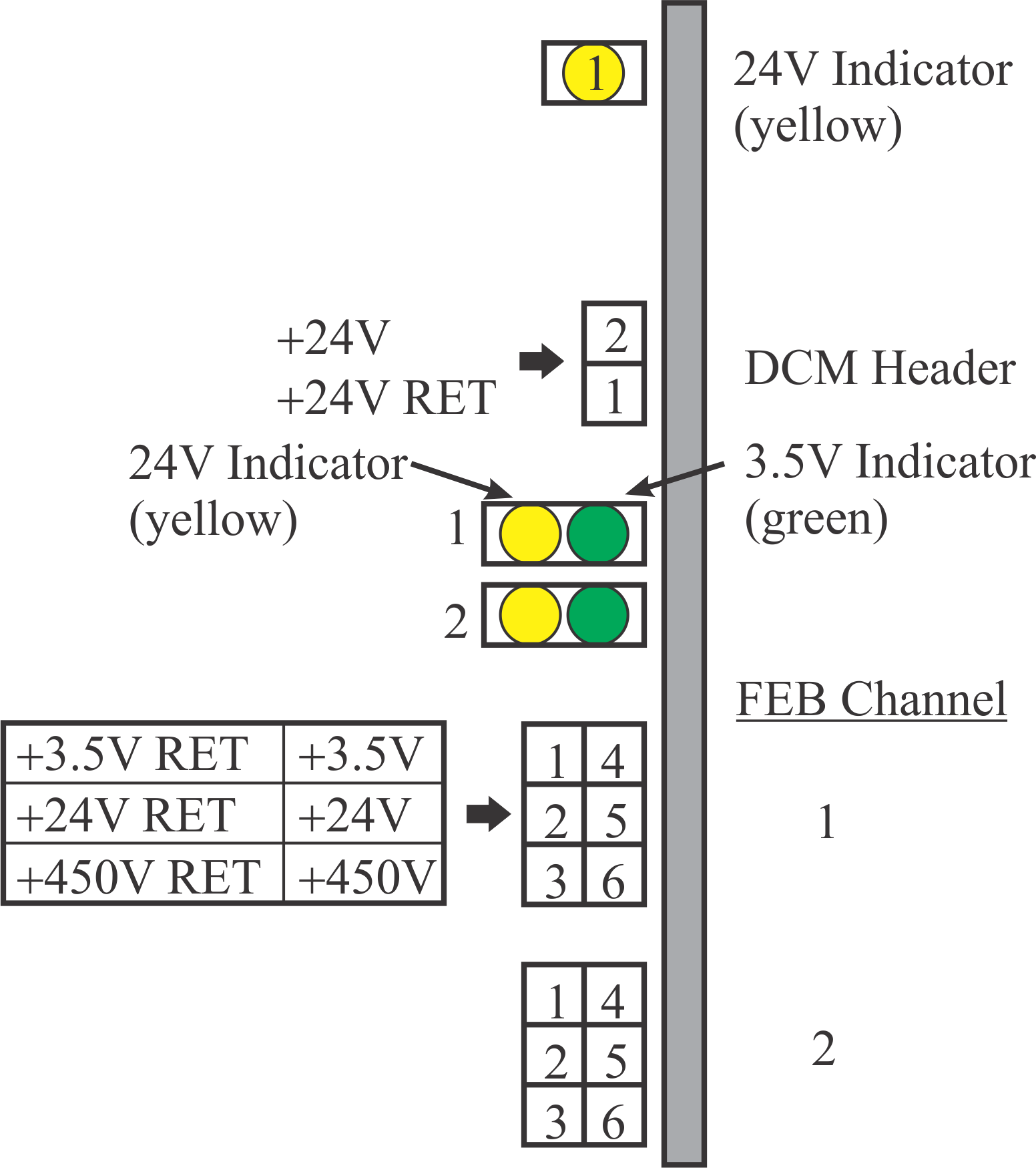}}
\caption[Front panel layout of power distribution box DCM card]%
{Front panel layout of power distribution box DCM card.  It feeds 24\,V to the data concentrator module and has two spare channels for FEB 
power.}
\label{fig:pdb_layout_channel_dcm}
\end{minipage}
\end{figure}

\begin{figure}[htbp]
\centering
\begin{minipage}[t]{6.0in}
\centerline{\includegraphics[type=jpg,ext=.jpg,read=.jpg,width=2.8in]{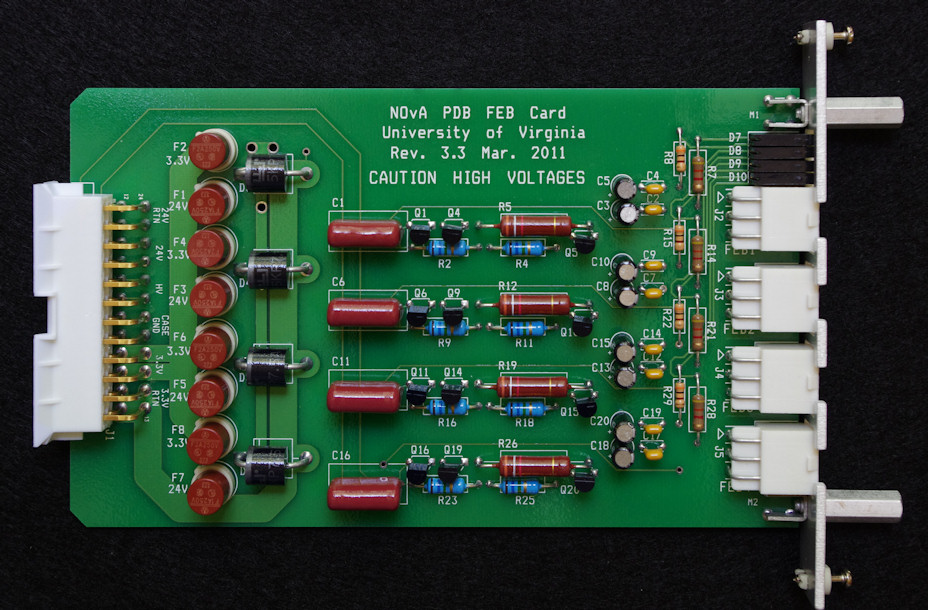}
\hspace*{\fill}
\includegraphics[type=jpg,ext=.jpg,read=.jpg,width=2.8in]{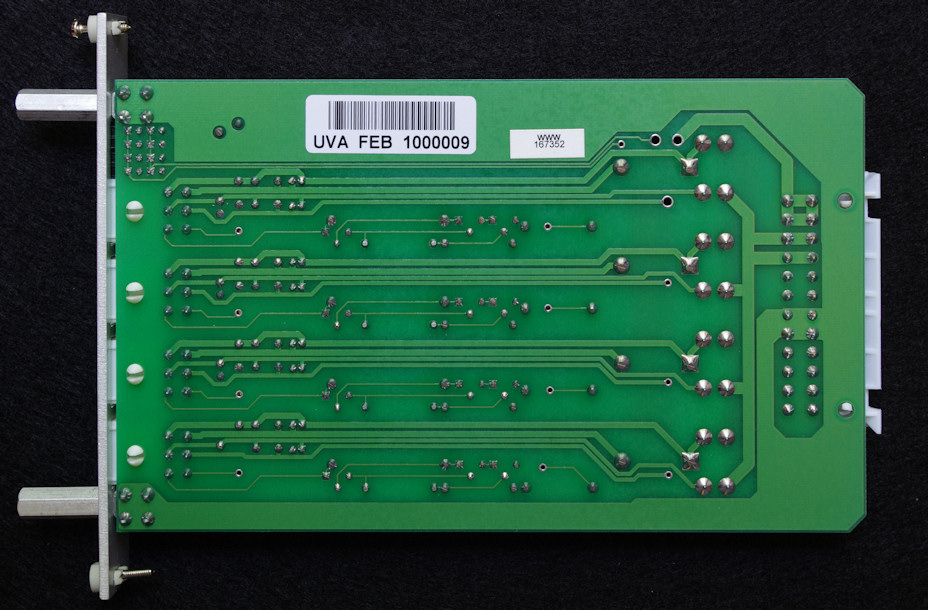}}
\caption[Power distribution box card that powers four FEBs]%
{Production version (V3.3) of the power distribution box card (FEB card) that powers four FEBs.}
\label{fig:card_feb}
\end{minipage}
\end{figure}
\hspace*{0.1in}%
\begin{figure}[htbp]
\centering
\begin{minipage}[t]{6.0in}
\centerline{\includegraphics[type=jpg,ext=.jpg,read=.jpg,width=2.8in]{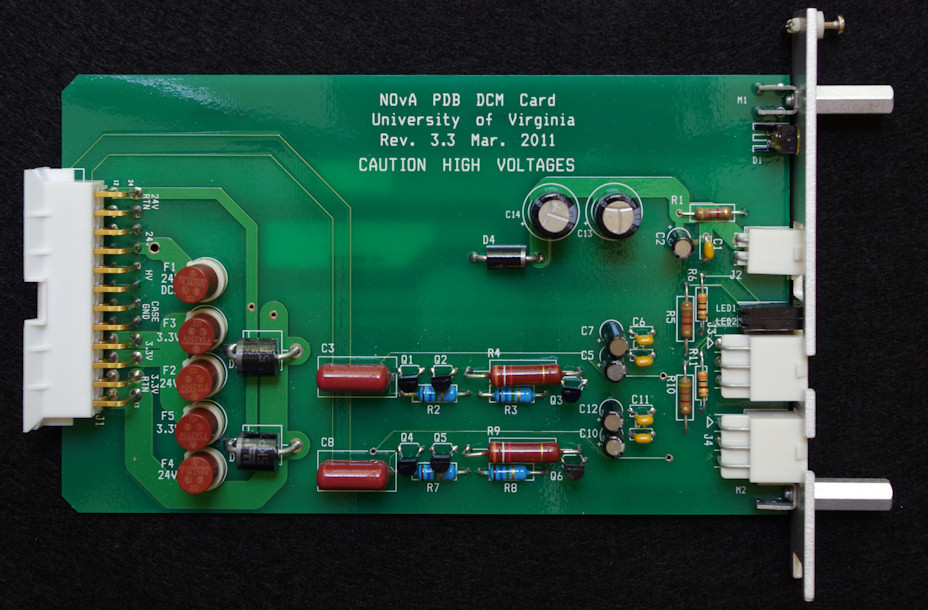}
\hspace*{\fill}
\includegraphics[type=jpg,ext=.jpg,read=.jpg,width=2.8in]{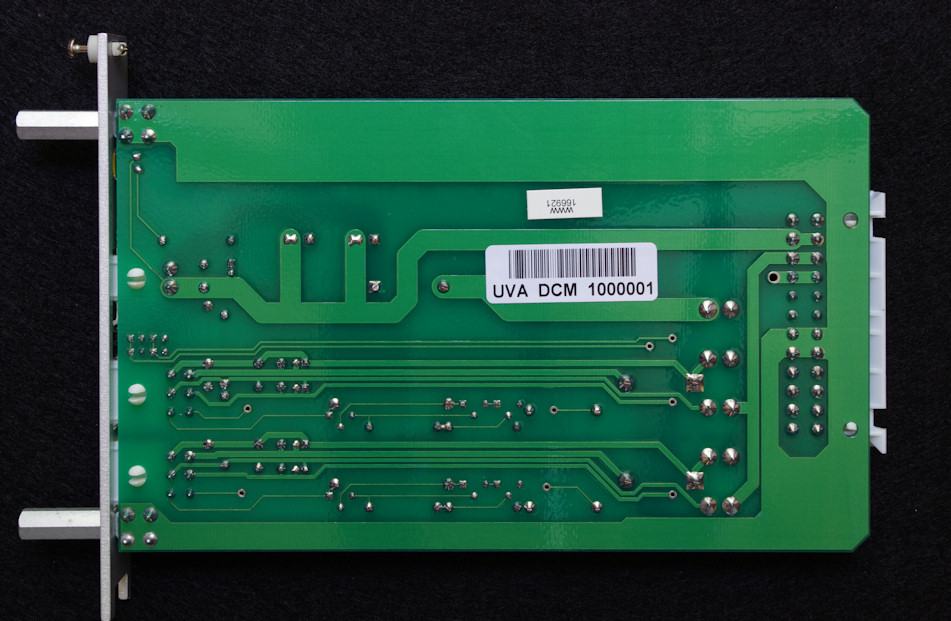}}
\caption[Card that powers a data concentrator module]%
{Production version (V3.3) of the power distribution box card (DCM card) that powers a data concentrator module (as well as has two FEBs.)}
\label{fig:card_dcm}
\end{minipage}
\end{figure}
\begin{figure}[htbp]
\centering
\begin{minipage}[t]{6.0in}
\centerline{\includegraphics[type=jpg,ext=.jpg,read=.jpg,width=2.8in]{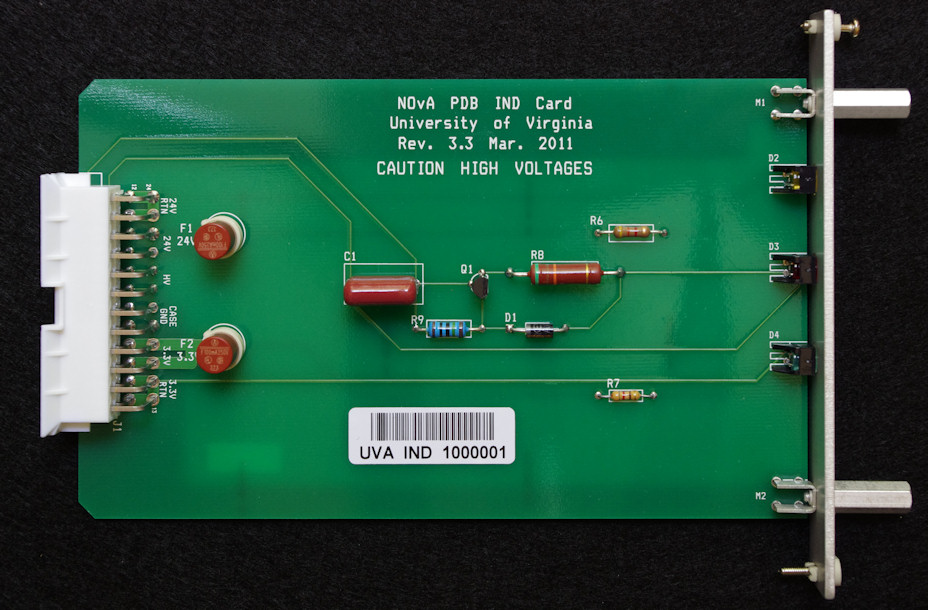}
\hspace*{\fill}
\includegraphics[type=jpg,ext=.jpg,read=.jpg,width=2.8in]{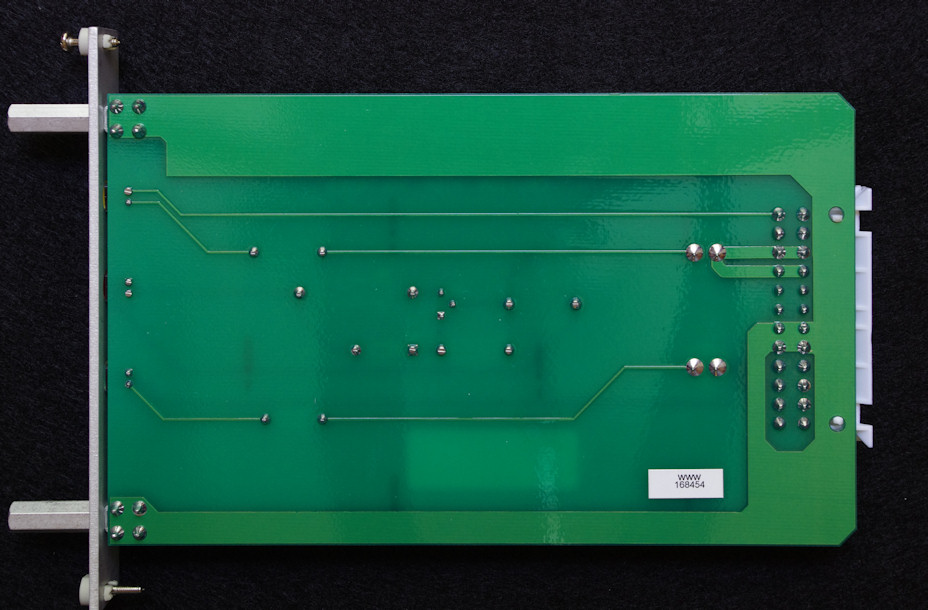}}
\caption[Power distribution box card with indicator lamps]%
{Production version (V3.3) of the power distribution box card (Indicator card) with indicator lamps.}
\label{fig:card_indicator}
\end{minipage}
\end{figure}

\subsection{FEB Card}

Each FEB card provides power to 4 front-end boards, allowing a maximum of 64 front-end boards 
to be powered through one power distribution box by the 16 FEB cards.  
The outputs are connected to the front-end boards via 6-conductor 18\,AWG cables using
6-pin receptacles \cite{header-6}.  The receptacles are keyed to prevent incorrect insertions.
Note that these receptacles are only rated for 20 insertions over their lifetime: receptacles
of the same size but rated to more insertions could not be found.
This has not been a problem as to date none have failed.
Two more FEBs can be powered by extra channels in the DCM card described below,
allowing a maximum of 66 FEB channels to be served by a single power distribution box.
Each 3.5\,V and 24\,V output is individually fused and has LED indicators.
There is no LED indicator for the high-voltage outputs.\footnote{A self-powered high voltage 
LED indicator would require more power than the high-voltage supply provides.}
Transient voltage suppressors on the 3.5\,V lines protect the
FEBs from over voltage due to an out of regulation power supply.

A current limiter for the high voltage lines limits the current to below 1.0~mA for safety reasons.
It has the added feature that a short in one front-end board, rather than
tripping the high-voltage power supply and disabling all of the high voltage channels served by the power distribution box,
only disables the shorted channel.   However, the limiter does produce a
current-dependent drop in the output voltage.  Because of this, and because
we wanted channel-by-channel bias control for individual APDs, a shunt
regulator voltage control circuit was added to the front-end board design
\cite{current_limiter}.

\subsection{DCM Card}

A special card, called the DCM card, feeds 24\,V power to the associated data concentrator module
and has two spare FEB power outlets.  The power to the DCM card is tapped off
before the 24\,V breaker, allowing the data concentrator module to be powered on while the 24\,V to
the TECs has been switched off using the front-panel breaker.\footnote{A front-panel switch for the 24\,V on the DCM card was desired, 
but a small enough one could not be found.}
An LED lamp indicates that the power is on.
The power distribution box crate and DCM card are keyed so that it can only fit in a special slot and no other.

\subsection{Indicator Card}

A third card, called the Indicator card, has LEDs that show which power supply voltages are 
being supplied to the power distribution box: 3.5\,V, 24\,V, and 425\,V.

\subsection{Testing}

A detailed description of the power distribution box tests is given in Ref.~\cite{testing-pdb}.
The tests were done using a jig designed and fabricated at the University of Virginia.
The testing was carried out using LabVIEW 2009 running on Windows XP on a National Instruments PXIe-1062Q chassis \cite{ni}.
Two power supplies provided three different voltages (6\,V, 30\,V, and 445\,V) to the 
power distribution boxes and three programmable electronic loads were used to put the DCM and FEB cards under load.
The testing was split into two parts: one LabVIEW program tested the power distribution box backplanes and
a second LabVIEW program tested the FEB, DCM, and Indicator cards.  
Relays allowed each channel and each slot to be tested under load, 
with the current and voltage data written to a database.
The backplane test was performed using special load boards plugged into the power distribution box backplane.
A standard power distribution box was used to perform this test.  
Another test checked the correct functioning of the high voltage current limiter circuit.
Cards that failed these tests were sent back to the manufacturer to be repaired.

\section{Cabling}

The power distribution system cables are listed in Table~\ref{tab:cable_types}.  
There are three power cables that run from the power supplies to the 
power distribution boxes on the detector: 3.5\,V, 24\,V, and 425\,V.  
As mentioned above, the 3.5\,V and 24\,V hot and return cables were twisted to reduce self induction
using a jig designed to fabricate rope \cite{rope}.  This eliminated a loud
noise when using the power supply remote sense function.
The 3.5\,V, 24\,V, and 425\,V power from the power distribution boxes to the front-end boards is carried 
by a single six-conductor cable and a single two-conductor cable carries the 24\,V power to the data concentrator modules.  

\begin{table}[htbp]
\footnotesize
\centering
\caption{\nova\ cable types.} 
\label{tab:cable_types}
\vspace*{0.08in}
\begin{tabular}{lccl}
  \toprule
  \multicolumn{1}{c}{Item}             & Cond. & AWG & Type \\
\midrule
  Power supply to PDB 3.5\,V           &     2 &   2 & Tray cable \\
  Power supply to PDB  24\,V           &     2 &   6 & Tray cable \\
  Power supply to breakout box 425\,V  &    37 &  26/7 & Multiwire HV cable \\
  Breakout box to PDB 425\,V           &     2 &  22 & RG-58A/U triaxial cable \\
  Sense                                &     4 &  22 & 2 pairs with shield and drain \\
  PDB to FEB                           &     6 &  18 & Tray cable \\
  PDB to DCM                           &     2 &  18 & Tray cable \\
  ground strip to PDB                  &   896 &  16\,mm$^2$ & $15{\times}1.5$\,mm$^2$ braid \\
  ground strip                         &     1 &  NA & $51{\times}0.4$\,mm$^2$ \\
\bottomrule
\end{tabular}
\end{table}

All cables were required to be rated either CMR (can be used in risers for commercial buildings) or CMP (can be used
in plenums) and all cables were flame-tested at Fermilab to insure they met internal safety requirements.
The cables used in the near detector had 38.1\,mm (1.5\inch) flame retardant shrink tubing placed over each
termination end for additional fire protection.

The power cables from the Wiener low voltage power supplies to the power distribution
boxes range in lengths from 6\,m to 21\,m.
Sense cables that run from the Wiener low voltage power supplies to the power distribution
boxes consist of two individually shielded twisted pairs.  
The sense lines are terminated on the power supply end to the manufacturer's pluggable
terminal strips.  At the power distribution box they are terminated in a
D-sub style DB-9 connector with an insulated housing.  The sense lines are attached
to the back of the power distribution box in a corresponding DB-9 receptacle.
The sense lines employ 0.50\,A inline fuses inside the power distribution box to prevent damage to the lines.  

The power cables from the Wiener high-voltage power supplies are routed to
a breakout box mounted on the same relay rack.  The cables running from the breakout box
to the power distribution boxes are triaxial cable.  
They range in lengths from 8\,m to 39\,m.

The power cables going from the power distribution boxes (PDB) to the FEBs are 6-conductor, 18\,AWG, 
non-paired, unshielded, tray cable, with a PVC jacket (Belden 27600A \cite{beldon}).
They are placed in wire basket cable trays of 152.4$\times$50.8\,mm$^2$ ($6{\times}2$\,in$^2$) size.
The PDB-FEB cables are grouped into four cable harnesses, each with 16 cables,
and of four different bundle types.  
They range in lengths from 1.2\,m to 3.8\,m.
A total of 672 (30) cable harnesses are used for the
far (near) detector (where we have excluded the near detector muon catcher from the total).
An amount of 10,752 (631) 6-conductor, 18\,AWG, cables carry the power from the power distribution 
boxes to the FEBs for the far (near) detector. 
The power supply to power distribution box cables and power distribution box to FEB cables were 
fabricated in Charlottesville, VA, USA and shipped out to Fermilab and far detector site \cite{cables}.

\section{Relay Racks}

The power supplies are mounted in deep 44U high relay racks situated on the upper catwalk at the
far detector and in an alcove at the near detector.  There is one relay rack per diblock
for the far detector low voltage supplies and two relay racks for the high voltage supplies
that serve the entire far detector.
For the near detector there are three relay racks, one with three and one with two low 
voltage power supplies.  The remaining near detector relay rack mounts the sole high-voltage power supply.
Figure~\ref{fig:rack_population_fd} shows a front view of the rack population for the far detector
low-voltage and high-voltage relay racks.

\begin{figure}[htbp]
\centerline{\includegraphics[width=2.5in]{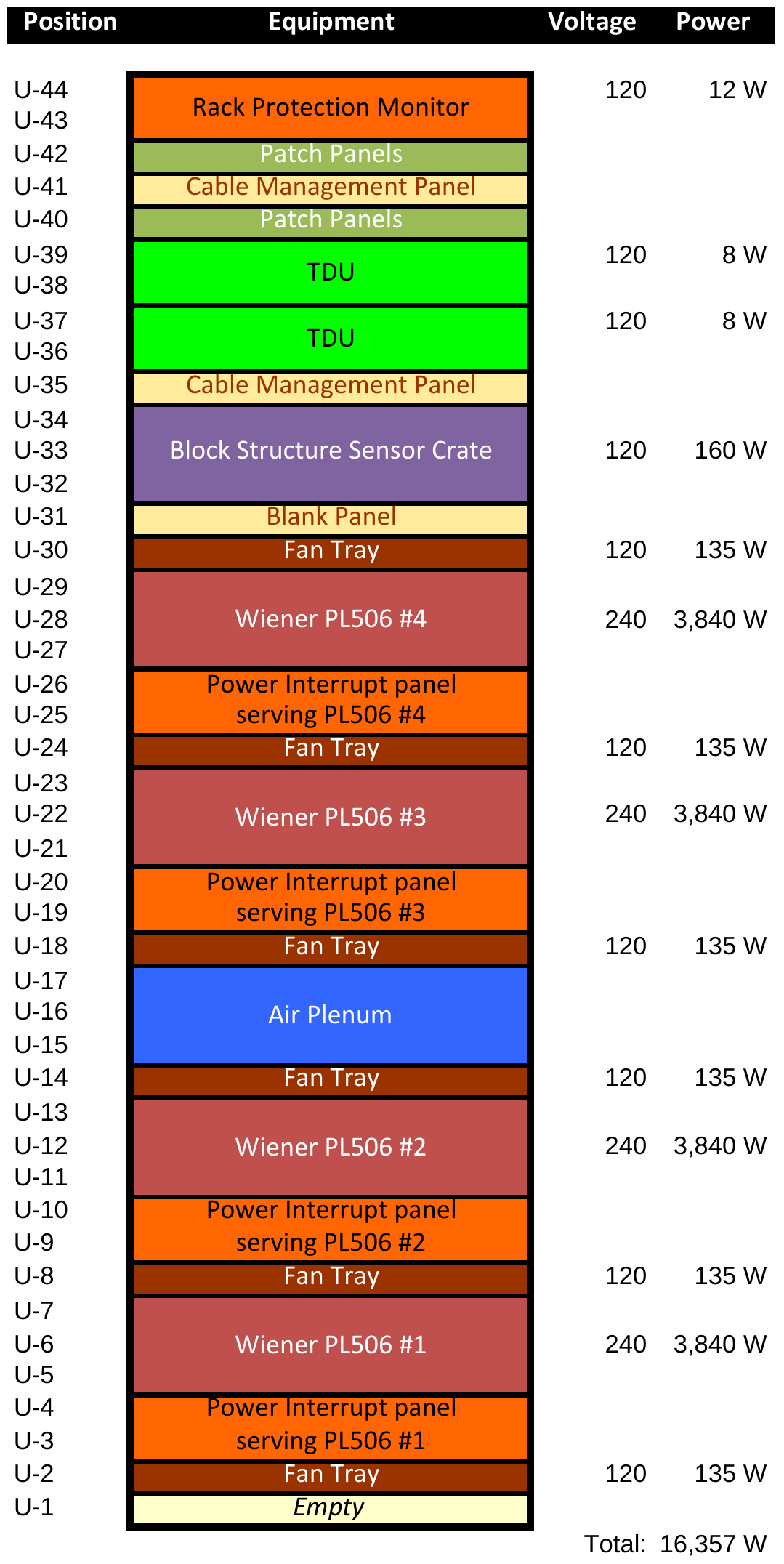}
\includegraphics[width=2.5in]{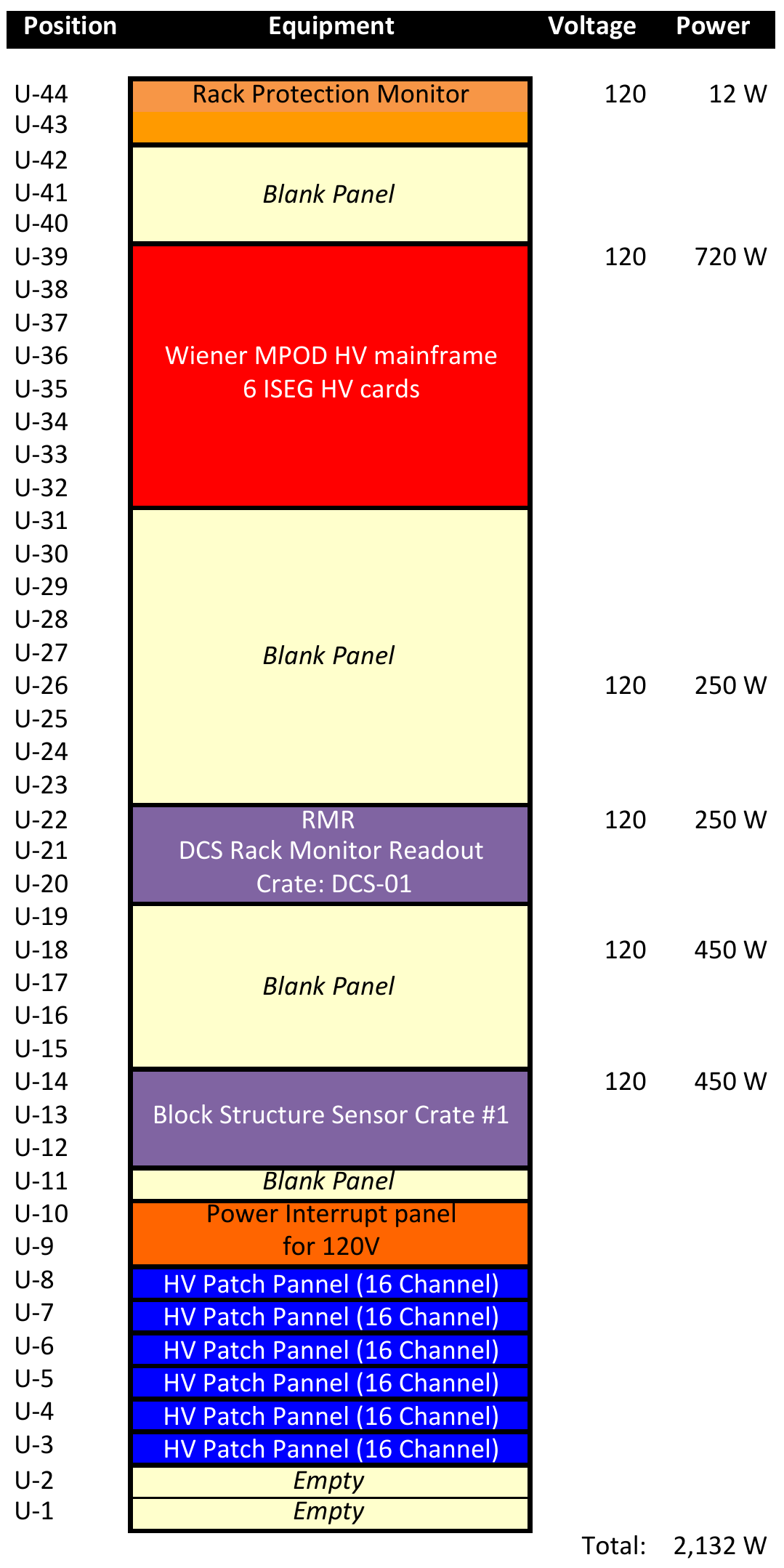}}
\caption[Far detector low-voltage relay rack population]%
{Left (right): far detector low-voltage (high-voltage) relay rack population.
}
\label{fig:rack_population_fd}
\end{figure}

The relay racks have a custom-made power interruption system \cite{rack_power} required for safety purposes.
It disables AC power to the equipment in the rack when smoke is detected, 
with the exception of the rack monitor itself and the field point system used to monitor the integrity of the detector structure.
Test results of the system are found in Ref.~\cite{rack_smoke_test}.

The proper functioning of the rack protection system imposes limitations for the airflow within the rack.
Specifically, the rack must form a chimney to funnel smoke to the sensors on the top.
The Wiener PL506 low voltage supplies consume quite a bit of power.  Their air intakes are
located below the units, with exhaust above.  A set of internal fans cool the units.  A test
was made with a single supply run at full power and outfitted with temperature sensors \cite{rack_cooling}.
The maximum temperature was found to be 48\tc, which is within the specified operational range of
0--70\tc\ for the supply.  However, to increase the time between failures for the power supplies we 
specified a temperature limit of 30\tc\ for each of the low-voltage power supply pods.
Additional cooling is provided by nine-fan, 1-U rack packs at the bottom of the relay racks
and above each power supply, giving a total of five (three) fan packs for the far (near) detector relay racks.
In addition a custom-designed plenum \cite{plenum} was installed between the second and third power supplies.
It brings in air from the front of the relay racks and deflects to the rear the warm air that is
blown upwards from the bottom two power supplies.
Typical average far detector low-voltage pod temperatures are 24\tc, 27\tc, 26\tc, and 28\tc,
going from bottom to top of the relay rack.  
Typical average pod temperatures for the near detector low-voltage power supplies are
24\tc, 28\tc, and 24\tc\ going from bottom to top in relay rack 2, and 
24\tc\ and 24\tc\ for relay rack 1.
Far (near) detector average high-voltage power supply temperatures at 27\tc\ (32\tc).

\section{Ground Scheme}

Great care was taken in the design of the far and near detector grounding schemes.
In particular, from the outset the design of the far detector building incorporated elements to insure a robust ground,
not only from a safety viewpoint, but to minimize high-frequency noise.
All transformers are single Faraday shielded units. All detector electronics have their own
transformers and these transformers are not shared with any other equipment.

\subsection{Far Detector Grounding Scheme}

The large size of the far detector, wide distribution of the electronics, and lack of large metallic surfaces
demand a good instrument ground.
Hence the building itself was designed to facilitate a high quality ground with little additional cost.
This method was pioneered by H.G.\ Ufer for ammunition storage facilities for the US military during World War II \cite{ufer}.  
Rebar is used rather than Ufer's copper wire, which has been shown to have resistances to earth ground as low as a 
fraction of an ohm \cite{fagan}.

The NOvA far detector building is large (100.8\,m [350\,ft] long, 20.16\,m [70\,ft] wide and 20.16\,m [70\,ft] high),
has ample rebar (\#6 rebar on 177.8\,mm [7\,in]  centers) and is semi-underground.
The catwalks are made of steel that is tied to the rebar every meter and is welded together to provide 
a continuous conducting structure. 
This large structure combined with multiple connections to the rebar provides the low inductance necessary for a good instrument ground. 
In addition, the spread out structure of the catwalk provides multiple current paths 
so magnetic coupling from ground current flows are minimized. 
Ground attachment points are placed on the catwalk next to the electronics racks.

The far detector grounding scheme is shown for a top and side power distribution box station in Fig.~\ref{fig:gnd_scheme_fd}.
Thick copper cables (2\,AWG) connect to thin (0.53\,mm [0.021\,in]) wide (50.8\,mm [2\,in]) copper ground strips that are placed
on the top of the detector, spaced apart by approximately 1.2\,m (4\,ft).
The large aspect ratio of the strips keeps their self inductance as low as possible.
The power distribution box enclosures and backplane ground busbars are connected to the ground strips by short 
ground braid strips \cite{braid}.
The shield of the high voltage triaxial cable, as well as the return, is connected to the ground braid,
as are the 3.5\,V and 24\,V return lines and sense cable shield, as shown in Fig.~\ref{fig:pds_schematic}.

The data concentrator module chassis sits on the same aluminum tray as the power distribution box; however the data concentrator module ground, unlike the 
power distribution box ground, is not directly connected to the data concentrator module chassis.

Cable trays on the top of the detector are connected by 6\,AWG cables to the ground strips.
On the side of the detector they are attached to Unistrut supports which are fixed to the
catwalk.

\begin{figure}[htbp]
\centerline{\includegraphics[width=5.0in]%
{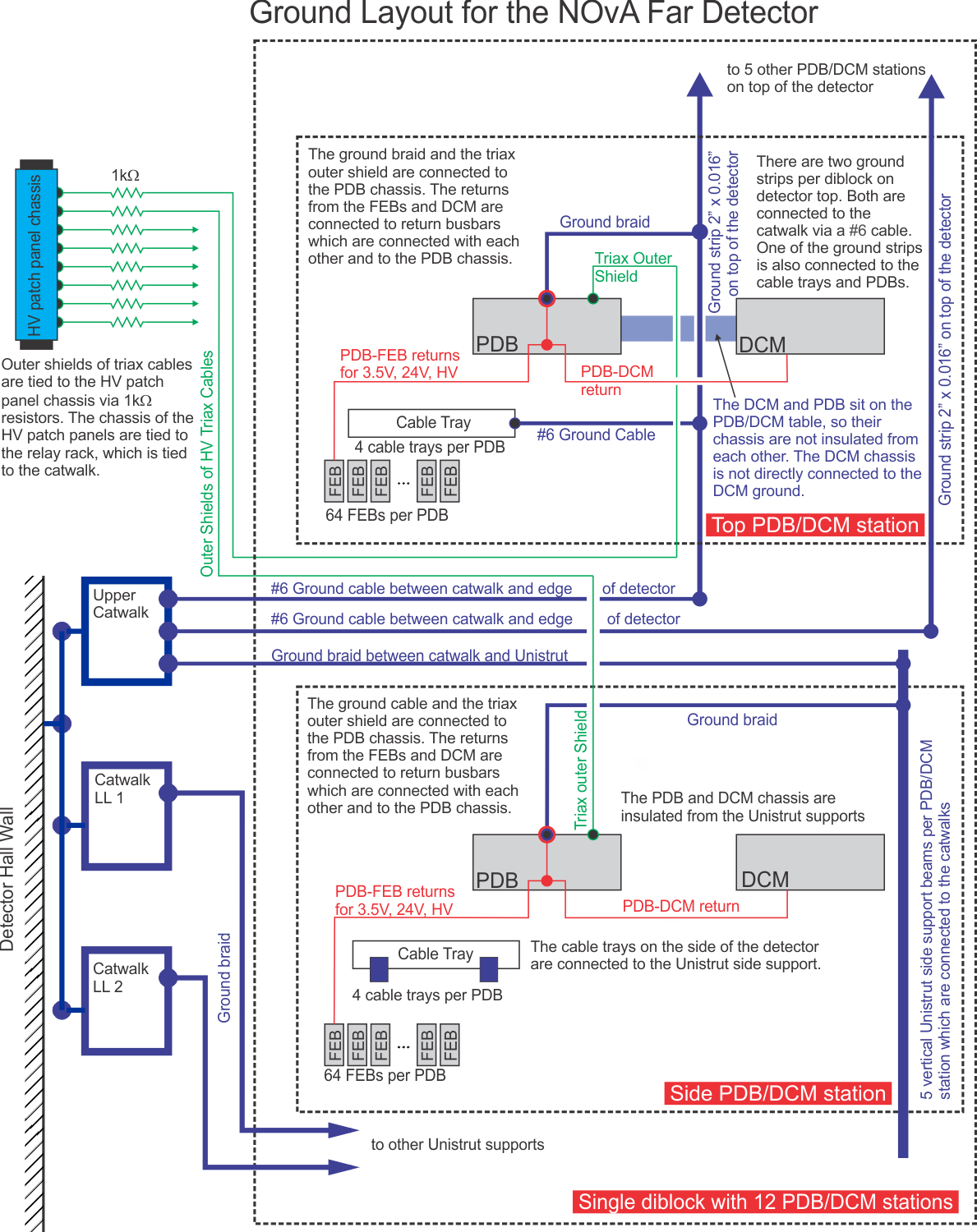}}
\caption[Far detector grounding scheme]%
{Far detector grounding scheme for one diblock.
}
\label{fig:gnd_scheme_fd}
\end{figure}

\subsection{Near Detector Grounding Scheme}

The NOvA near detector ground scheme, shown in Fig.~\ref{fig:gnd_scheme_nd}, is similar to that
of the far detector.  
Noise issues are mitigated by the small size of the detector and its location over 100\,m underground.
Also, on average the signals are larger due to the shorter fiber length.

\begin{figure}[htbp]
\centerline{\includegraphics[width=5.0in]%
{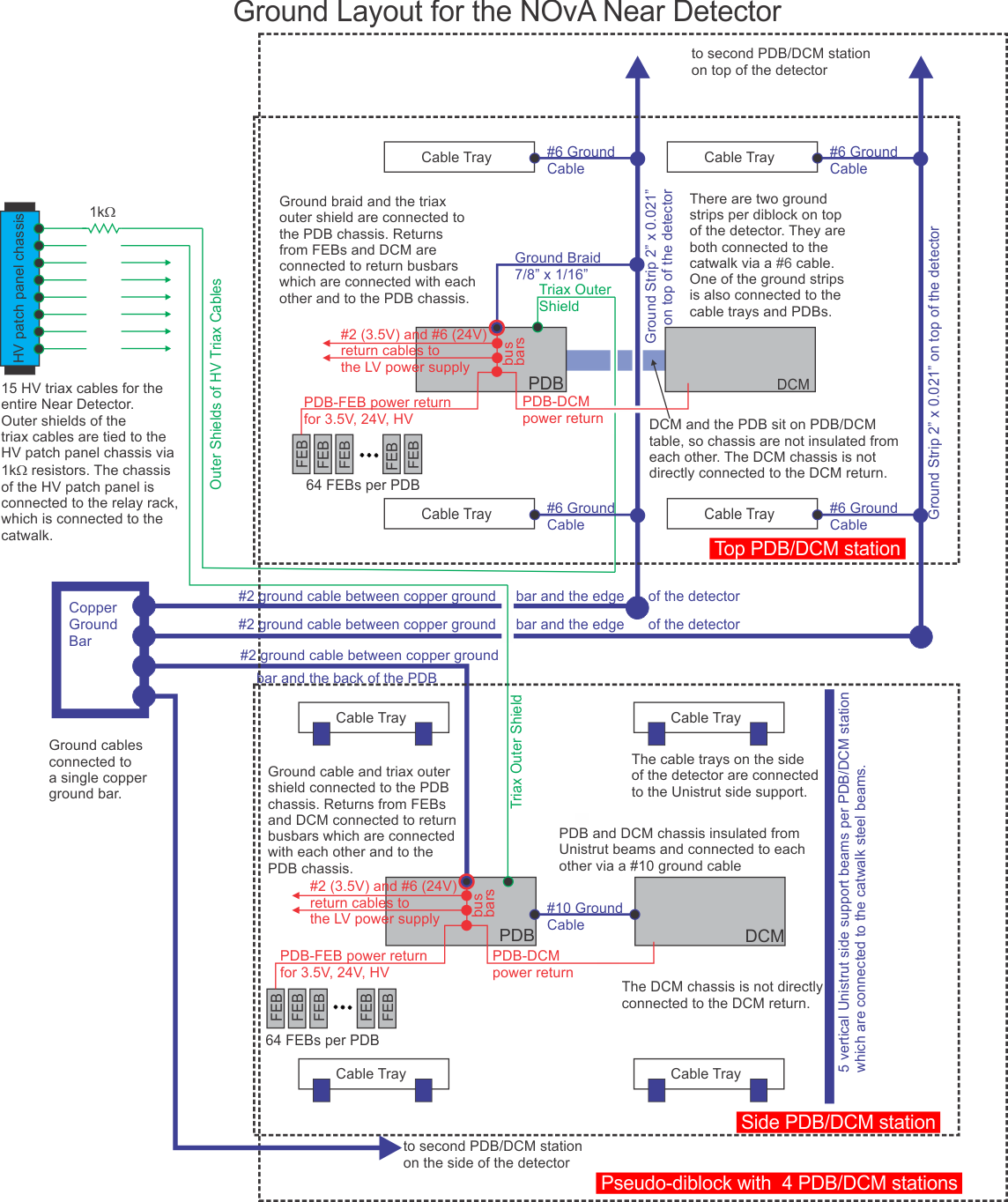}}
\caption[Far detector grounding scheme]%
{Near detector grounding scheme.
}
\label{fig:gnd_scheme_nd}
\end{figure}

\section{Operation}

The power distribution system has worked extremely well in the four years of detector operation
with very little loss of detector livetime due to any failures.  
No high-voltage power supply channels have failed.
There have been eight low-voltage power supply failures.
They include: a tripping power supply pod, an unresponsive power supply,
non-functioning temperature sensors, and pods which reported inaccurate sense channels.
Failed power supplies have been immediately replaced by spares.
The only other failures to the system have been several power distribution boxes that were damaged by
water leaks in the avalanche photodiode cooling system at the beginning of the data taking.
They were replaced.

Typical operating voltages (at the power distribution boxes) and currents for
the front-end boards and thermoelectric coolers are given in Table~\ref{tab:currents}.
Note the higher maximum front-end board currents in the near detector due to the higher digitization rates.
\begin{table}[htbp]
\footnotesize
\centering
\caption[Voltages and Currents]{Range of near and far detector power distribution box voltages and currents.}
  \label{tab:currents}
\vspace*{0.08in}
\begin{tabular}{lcccc}
  \toprule
  & \multicolumn{2}{c}{Near Detector} & \multicolumn{2}{c}{Far Detector} \\
  Component & \multicolumn{1}{c}{Terminal Voltage} &  \multicolumn{1}{c}{Current} & \multicolumn{1}{c}{Terminal Voltage} 
  & \multicolumn{1}{c}{Current} \\
\midrule
  FEB (3.5\,V)   &   4.1--5.2\,V & 23.0--66.6\,A  &  3.9--4.4\,V & 30.3--30.6\,A \\
  TEC (24\,V)    &  24.2--24.7\,V &  4.4--12.1\,A  & 24.2--24.6\,V &  8.2--9.7\,A \\
  APD (425\,V)   &     425\,V     &  2.2--3.9\,mA &    425\,V     &  3.2--4.2\,mA \\
\bottomrule
\end{tabular}
\end{table}

\section{Acknowledgments}

We wish to acknowledge the assistance of J. Oliver of Harvard University in the design of the high-voltage limiting circuit,
M. Johnson of Fermilab in the design of the grounding system, and M. Matulik in the design of the relay rack protection system.
We acknowledge the important contributions to the design, fabrication, and testing of the \nova\ power distribution
system from University of Virginia undergraduate students: G. Bailey, D. Evans, N. Fields, J. Gran, E. Ho, 
S. Hasselquist, K. Lagergren, D. Mudd, S. Russo, H. Tammaro, and K. Tran; technicians B. Mason and  L. St.\ John; 
and graduate student Z. Wang.
This work was supported by the US Department of Energy. 
Fermilab is operated by Fermi Research Alliance, LLC under Contract No.\ DE-AC02-07CH11359 with the US DOE.

\clearpage



\end{document}